\documentclass[12pt,a4paper,oneside]{article}

\usepackage[main=english]{babel} 
\usepackage[utf8]{inputenc} 
\usepackage[T1]{fontenc} 
\usepackage{mdwlist}
\usepackage{enumitem}
\usepackage{longtable}
\usepackage{physics}
\usepackage{hyperref}
\usepackage{array}
\usepackage{graphicx}
\usepackage{amssymb}
\usepackage{amsmath}
\usepackage{amsthm}
\usepackage{ifthen}
\usepackage{fancyhdr} 
\usepackage{caption}

\newcommand{\comment}[1]{ }  

\comment{
\newlength\tindent
\setlength{\tindent}{\parindent}
\setlength{\parindent}{0pt}
\renewcommand{\indent}{\hspace*{\tindent}}
}

\comment{
\let\oldtabular=\tabular
\def\tabular{\footnotesize\oldtabular} 
}

\graphicspath{{./}}

\newlength{\pagewidth}
\newcommand{\typeRendu}{\textit{Report}}
\newcommand{\titre}{Study of the effects of incentive measures on Covid-19 vaccination in the United States}
\newcommand{\dateRendu}{05/30/2022}
\newcommand{\shorttitle}{Mathematical Modeling of Complex Systems}

\begin{document}

\fancyhead[RO]{\fancyplain{}{\shorttitle}}
\fancyhead[LO]{\fancyplain{}{\typeRendu}}

\pagenumbering{arabic}

\begin{titlepage}
\thispagestyle{empty}
\vspace{-1.5cm}
\includegraphics[height=1.5772cm]{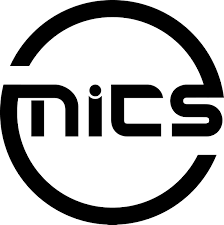} 
\hfill
\includegraphics[height=1.5772cm]{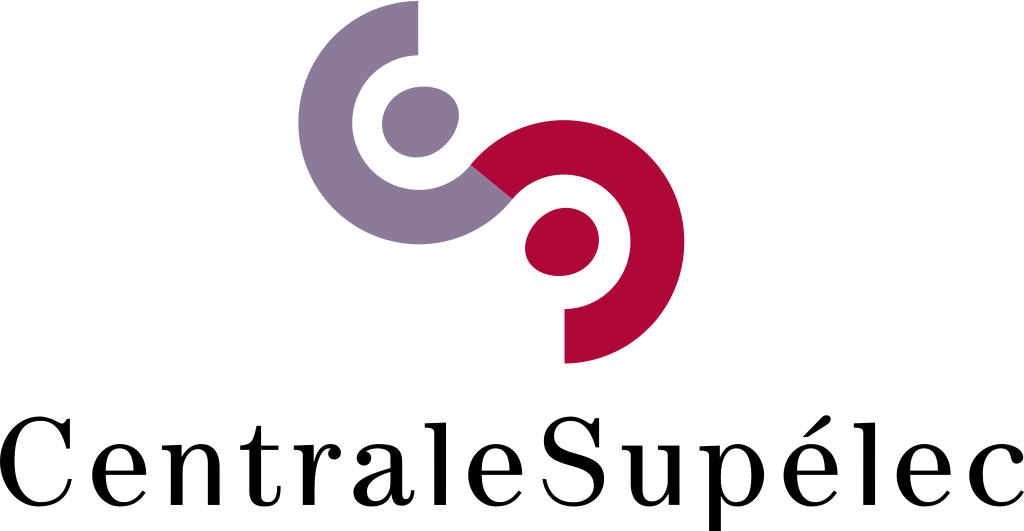}

\vfill

\begin{center}
\Huge{\titre}
\\
\vspace{0.2cm}
\large{\dateRendu}
\\
\vspace{0.2cm}
\large{\typeRendu}
\end{center}
\vspace{0.1cm}

\begin{center}
\includegraphics[width=10cm]{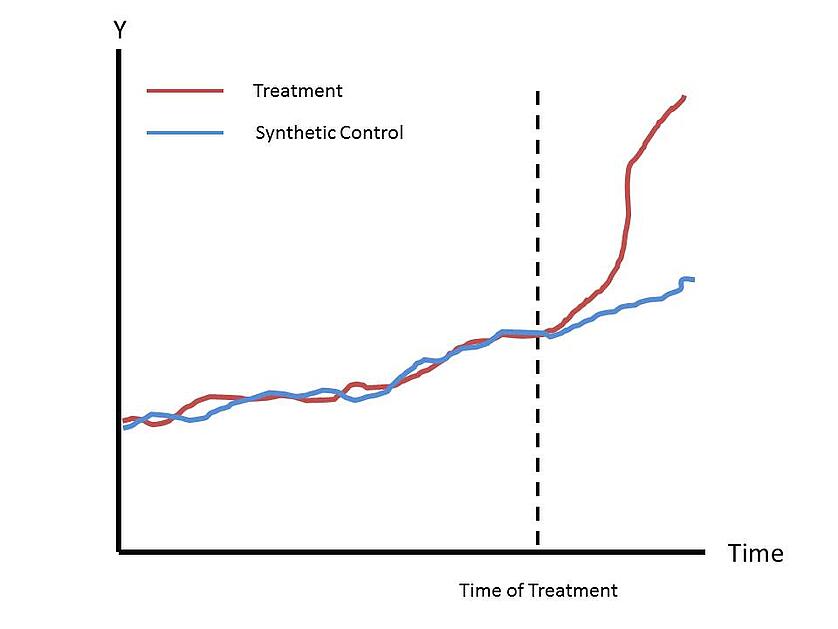}
\end{center}

\vfill
\begin{minipage}[t]{0.5\textwidth}
\centering
{\large \scshape TEAM MEMBERS :}\\
{Hector BONNEFOI}\\
{Théodore DE POMEREU}\\
{Lucas LEFORESTIER}\\
{Antoine LOPES FERREIRA}\\
{Vishnou VINAYAGAME}
\end{minipage}
\begin{minipage}[t]{0.5\textwidth}
\centering
{\large \scshape Client and referral:}\\
{Marie-Laure CHARPIGNON}\\
\end{minipage}
\vspace{1cm}\\
\rule{\textwidth}{1pt}
\begin{tabular}[t]{l}
CentraleSupélec \\
Mathematical Modeling of Complex Systems\\
\end{tabular}
\hfill
\begin{tabular}[t]{r}
\\
\\
\end{tabular}
\rule{\textwidth}{1pt}

\end{titlepage}

\cleardoublepage

\section*{Acknowledgements}

We would first like to thank the pioneers of the "Mathematical modeling of com- plex systems" cluster at CentraleSupelec: Ioane Muni-Toke, Véronique Letort-Le Chevalier and Gurvan Hermange.\\ \\ \indent
We wish to express our deepest gratitude to our referent for this project, Marie-Laure Charpignon. Your commitment at our side has been beyond all our expectations. Thanks to you, this "great project" of the last two years has made us grow academically, technically, professionally and of course humanly. We are very proud of the path we have taken and of the work we have accomplished, to the point of being able to say that this project has probably been the most enriching academic experience of our two years at CentraleSupélec. A big thank you for being a true "mentor", within the project and beyond!
\newpage
\section*{Abstract}

With COVID-19 having emerged as the most widespread human pandemic disease in a century, the need to control its spread to avoid massive loss of life became more than necessary, and extremely fast. Several vaccines were developed and the task of policy makers was suddenly to convince the reluctant population to be vaccinated by various means. While some countries have chosen a policy of mandatory vaccination or punitive incentives, many states in the United States have adopted various incentives to try to increase vaccination coverage. A study we conducted in recent months quantified the effect of these measures on the proportion of the population vaccinated, using the synthetic control method, by simulating what would have happened without these measures. The aim now is to generalize this study to smaller scales, to improve the results of our previous study, to quantify their robustness and to provide a tool that can be used by policy makers to adapt their behavior in light of the results obtained.

\indent 
\newpage
\cleardoublepage
\tableofcontents
\newpage

\section{State of the Art}
 
\subsection{Incentives for vaccination}
\indent In order to carry out our study, we had to collect and list the incentives that were put in place in the different states of the United States. We were thus able to classify these different measures into 4 groups according to the type of reward offered, while taking into account the fact that some measures were subject to lotteries and thus had a random element in the reception of the reward, or not. The 4 categories were thus established as follows:
\begin{itemize}
    \item \textbf{Category 1:} The smallest prizes, almost systematically not subject to a lottery, and including rewards in the order of ten dollars in cash or vouchers, drinks offered in restaurants, or free sports or cultural activities.
    \item \textbf{Category 2:} Intermediate prizes, often vouchers in the hundreds of dollars, free vacations, or tuition discounts (very high in the United States).
    \item \textbf{Category 3:} Large prizes, almost systematically submitted to lotteries with a very small number of winners (usually 1 to 10 statewide, depending on the prizes offered). The prizes often range from \$5,000 to several million dollars, or full scholarships over several years.
    \item \textbf{Category 0:} Unclassifiable or incomparable lots, often because they concern too small a part of the population (e.g. employees of a particular chain store, prisoners or civil servants)
\end{itemize}
\subsection{Synthetic Control}
\indent In order to model what would have happened in each target state in the absence of an incentive, we will again use the synthetic control method, developed by Alberto Abadie, whose mathematical approach is detailed below.

\subsubsection*{Mathematical Model}
We define:\newline
- $T$ : the total number of periods and $T_{0}$ the number of periods before the intervention \newline
- $j \in [1, J+1]$ : the total number of entities (here States) with $j = 1$ the entity undergoing the intervention \newline
- $Y_{jt}$, the result of interest for each entity j and time portion t \newline
- $X_{kj}$, for each entity we observe $k$ predictors of the outcome which include pre-intervention values and are not affected by the intervention. \newline
We then create the vectors $\textbf{X}_{j}$ of dimension $k$ x $1$ with $\textbf{X} = [\textbf{X}_{2}, \textbf{X}_{3}, ..., \textbf{X}_{J+1}]$, collecting information for the $J$ without intervention \newline
- $Y_{jt}^{N}$, potential result without intervention
\newline
- $Y_{1t}^{I}$ = $Y_{1t}$, for $t > T_{0}$ results after intervention \newline
- $\tau_{1t} = Y_{1t}^{I} - Y_{1t}^{N}$, the effect of the intervention, the goal being to determine $Y_{1t}^{N}$\newline
\\
We create the estimator of $Y_{jt}^{N}$ :\newline
\begin{equation}
    \hat{Y}_{jt}^{N} = \sum_{j=2}^{J+1} w_{j}Y_{jt}
\end{equation}
\begin{equation}
    \hat{\tau}_{1t} = Y_{1t}^{I} - \hat{Y}_{1t}^{N}
\end{equation}
with $\sum_{j=2}^{J+1} w_{j} = 1$ et $w_{j} \geqslant 0$
\newline

The question remains how to choose the weights. We choose $\textbf{W} = (w_{2}^{*}, .., w_{J+1}^{*})$ such that: 
\begin{equation}
    \parallel \textbf{X}_{1} - \textbf{X}\textbf{W} \parallel = (\sum_{h=1}^{k} v_{h}(X_{h1} - w_{2}X_{h2} -...  - w_{J+1}X_{hJ+1})^{2})^{1/2}
\end{equation}
is minimized, with $\textbf{V} = (v_{1}, ..., v_{k})$ non-negative coefficients that represent the importance of each predictor. Equation (3) is easily solvable using "constrained quadratic optimization".
\newline
\\
Then we must choose $\textbf{V}$ such that it minimizes MSPE (mean squared prediction error)
\begin{equation}
    \sum_{t \in ET_{0}} (Y_{1t} - w_{2}(\textbf{V})Y_{2t} - ... - w_{J+1}(\textbf{V})Y_{J+1t})^{2}
\end{equation}
With $ET_{0} \in {1, ..., T_{0}}$\newline
The choice of $\textbf{V}$ amounts to giving importance to each ${X}_{l1}$ as a predictor of $Y_{1t}^{N}$.

\subsubsection*{Principle of the method}

1. We divide the pre-intervention periods into an initial training period and a subsequent validation period. For simplicity and concreteness, we will assume that $T_0$ is even and that the training and validation periods span $t=1,...,t_0$ and $t=t_0 +1,...,T_0$ respectively, with $t_0 = T_0/2$. In practice, the duration of the training and validation periods may depend on application-specific factors, such as the extent of outcome data availability in the pre- and post-intervention periods, and the specific times at which predictors are measured in the data.
\newline
\\
\indent 2 . For each $\textbf{V}$, $w_{2}(V), ..., w_{J+1}(V)$ are the synthetic control weights computed with the training period data on the predictors. The MSPE of the synthetic control over $Y_{1t}^{N}$ for the validation period is :
\begin{equation}
    \sum_{t= t_0+1}^{T_{0}} (Y_{1t} - w_{2}(\textbf{V})Y_{2t} - ... - w_{J+1}(\textbf{V})Y_{J+1t})^{2}
\end{equation}
\\
\indent 3. We take $\textbf{V}^{*}$ such that the MSPE (5) is arbitrarily small
\\
\indent 4. We use the results of $\textbf{V}^{*}$ and the predictor data for the last $t_0$ periods before the intervention to find $\textbf{W}^{*}$.

\subsection{Continuity with precedent work}
\indent The work of modeling synthetic states in order to measure the real impact of incentive measures on the vaccination of populations in the states had been started by our group in September 2021 within the MICS Lab of CentraleSupélec. However, there were still many points to come back to, which could be improved, and other new axes to tackle, which was the objective of our tasks for this report. \newline
The first axis that we wanted to study, which we could not do in the previous project due to lack of time, was the robustness of the model. To do this, we chose to work on the unexplained differences, for some states, that we obtained between the real curve and the synthetic vaccination curve, over the pre-intervention period. This results in a larger number of states for which the fitting is good and gives interpretable results. \newline
\indent On the other hand, we regretted that there was too little measurement of the quality of the results obtained, which is why we were able to calculate and establish p-values on the modeling results of all the targeted states. \newline
\indent We also wanted to develop our understanding of the problem through an in-depth study of the demographics of the populations concerned, by bringing in new objects of study such as the community vulnerability index. \newline
\indent The continuity of this project with the previous one is mainly due to our wish to generalize the model, especially by counties (territorial subdivision of a state, governed by a local government and grouping several municipalities).\newline
\indent Finally, our previous work included a dashboard summarizing our results and acting as an appendix to our report, created with Dashly. We wanted a total redesign of this interface, both graphically and structurally. Our objective was to create an interactive tool that could be considered independently of this report, and serve an informative purpose for public health decision makers.
 
 \newpage

\section{Improvement of the statewide model}
A first part of our work during this semester was to revisit our results from the S7 project. We wanted to go back to some inaccuracies of our previous model and explore ways to improve it. The goal is first to improve the robustness of the model and then to improve the quality and reliability of our results.
The first four sub-sections that follow report on our work to improve the robustness and quality of the model. The last two sections concern the p-values and the results of the model.

\subsection{Selection of parameters}

During the first part of the project at the beginning of the year, we performed the synthetic control method on all the American states that had incentives. However, in the case of some states, the fitting over the learning period before the intervention period sometimes showed too high a difference between the synthetic curve and the real curve, which resulted in inconsistencies afterwards. One of the solutions was therefore to reduce the large number of parameters (initially 30) that characterized our states.

\subsubsection{Naïve Method}

To remedy this, we chose to try a naive method by grouping the parameters into 6 distinct blocks, corresponding to themes (demographic, ethnic, education, economic, political and health), and by selecting 2 parameters within each of them, which gives us 12 parameters in total. For the selection of the parameters within the same block, we initially choose the parameter with the highest correlation with the others, then for the second one, with the remaining parameters, we remove all those with a too high correlation with the first one chosen by defining an arbitrary threshold (here 0.4), and we choose the one with the highest correlation among the remaining ones (in the same way as for the choice of the first).

\begin{center}
    \includegraphics[width= 12cm]{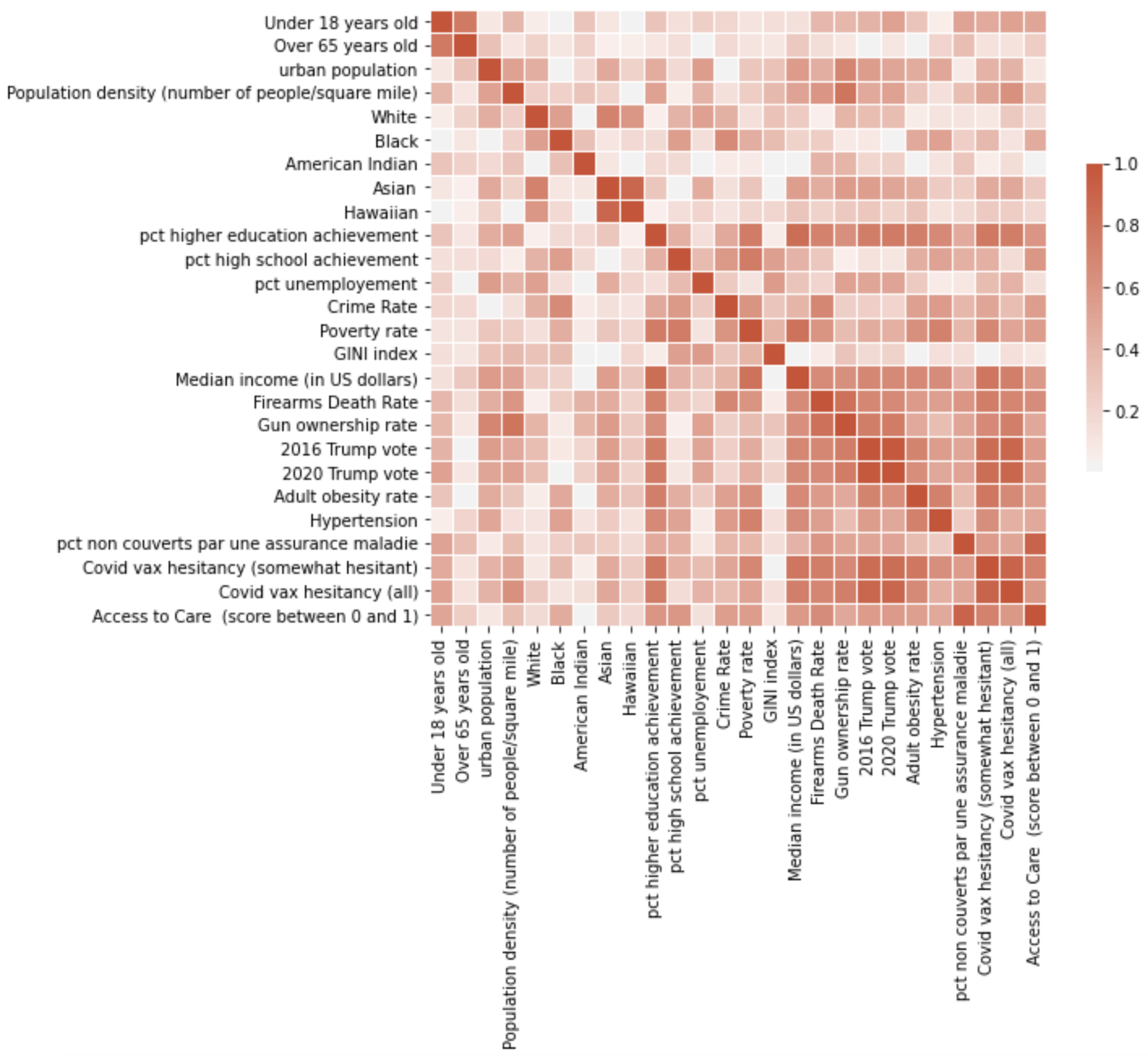}
\end{center}
\begin{center}
    \textit{Figure 1: Correlation matrix (in absolute value) of the different parameters}
\end{center}

Thus, after our selection, the list of 12 parameters is as follows: Population under 18, Proportion of white people, Percentage having studied in higher education, Gun death rate, Vote for Trump in 2016, Covid vaccination hesitation, Urban population, Proportion of Native American people, Percentage having studied in high school, Unemployment rate, Vote for Trump in 2020, and Morbid obesity rate in adults. 

\subsubsection{Results}

With regard to the results, given our naive approach to parameter selection, the conclusions must be qualified. Indeed, even if for some states, the selection allowed to drastically reduce the delta between the real curve and the synthetic curve over the whole fitting period, for example with Alabama by going from -40 to -0.7, sometimes the opposite effect is observed with the example of Connecticut, where the new synthetic curve with less parameters gives a delta of 166 against 9 for the initial synthetic curve. Thus, although the results seem to be better for the majority of states, it is important to emphasize the naive approach of the method, which can sometimes lead to large discrepancies, as shown above.

\subsection{Selection of variables}

One way to improve the model was to fight more efficiently against over-fitting which was a possible source of nuisance. To do so, we tried to remove the selection of variables ($v_1$, ..., $v_k$) from the optimization algorithm by fixing them a priori for all treated states. One method proposed in Abadie (2021), is to take each $v_i$ as equal to the inverse of the variance of the parameter it weights. This allows the fitting to focus on those parameters that most explain the differences between states.
The effect on the model was disappointing, however, as the fits over the pre-intervention period were of poorer quality than those of the original model.

\subsection{Training period}

Another question that was not well explored during the previous semester was the optimal choice of the training period. Computational time savings in several places in the notebook concerned allowed us to estimate more efficiently the evolution of the quality of the pre-intervention fitting as a function of the duration of the training period chosen. 
\\
\indent The date for which state-specific vaccine data begins to be reported for all states is approximately February 19, 2021. The incentives themselves begin in large part in May 2021 at the earliest. This leaves a little over 2 months or about 70 days of duration for the pre-intervention period. \\
\indent We therefore chose to evaluate training periods of between 10 and 50 days.\\
With respect to the validation period, there are two considerations specific to our study that are very important to take into account.
\begin{itemize}
\item First, the vaccination curves are extremely close in the pre-intervention period. It is around the period of implementation of the incentive that we can observe the beginning of the differentiation of the different states in their vaccination rates. This gives great importance to the 10-20 day window before the implementation of the incentive. The validation period is therefore not ideal in this context because it affects the quality of the fit.
\item Then, the quality of the fits can be roughly divided into three groups: "good" (there is a negligible discrepancy between the synthetic and real curves over the pre-intervention period), "average" (we notice small discrepancies over periods of less than ten days) and finally "bad" (there is a pronounced discrepancy between the real and synthetic curves over the majority of the pre-intervention period) We found that the vast majority of the fits were either "good" or "bad". Our priority is therefore to find for each state in which "category" of fitting quality each treated state falls. The quantity chosen to evaluate the quality of the fit was therefore the quadratic difference between the real and synthetic curve over the whole pre-intervention period, including the training period.\\
\end{itemize}

\indent The results of the fitting period analysis suggest that the length of the training period has little influence on the quality of the fit. However, we can notice that the choice of a training period between 10 and 20 days seems slightly better. This added to the slight gain in computation time that this choice induces, makes us choose 10 days as our reference training period in the rest of the study. 

\subsection{Regularization}

Regularization was an important addition to the first model done last semester. To save computational time and escape over-fitting by reducing the number of control states contributing to the synthetic curve, we studied the effect of elastic-net on our model.
The elastic-net is the combination of ridge and lasso regularization.  \\
\indent It results in the addition of two terms in the term to be minimized each weighted by a coefficient ($L_1$ and $L_2$) in the optimization sequence in the synthetic control:

$$\left(\sum_{h}^{} v_h ( X_{h}- \sum_{i}^{} w_i X_{hi})^2 \right)^{1/2}+L_1 \lVert \textbf{X} \rVert_2 +L_2 \lVert \textbf{X}\rVert_1$$
\indent In the table above,the effects of elastic-net as a function of the value of $L_1$ and $L_2$ are summarized . The measure in the table is the squared deviation between the synthetic and real curves during the pre-intervention period.

\begin{center}
    \includegraphics[width=16cm]{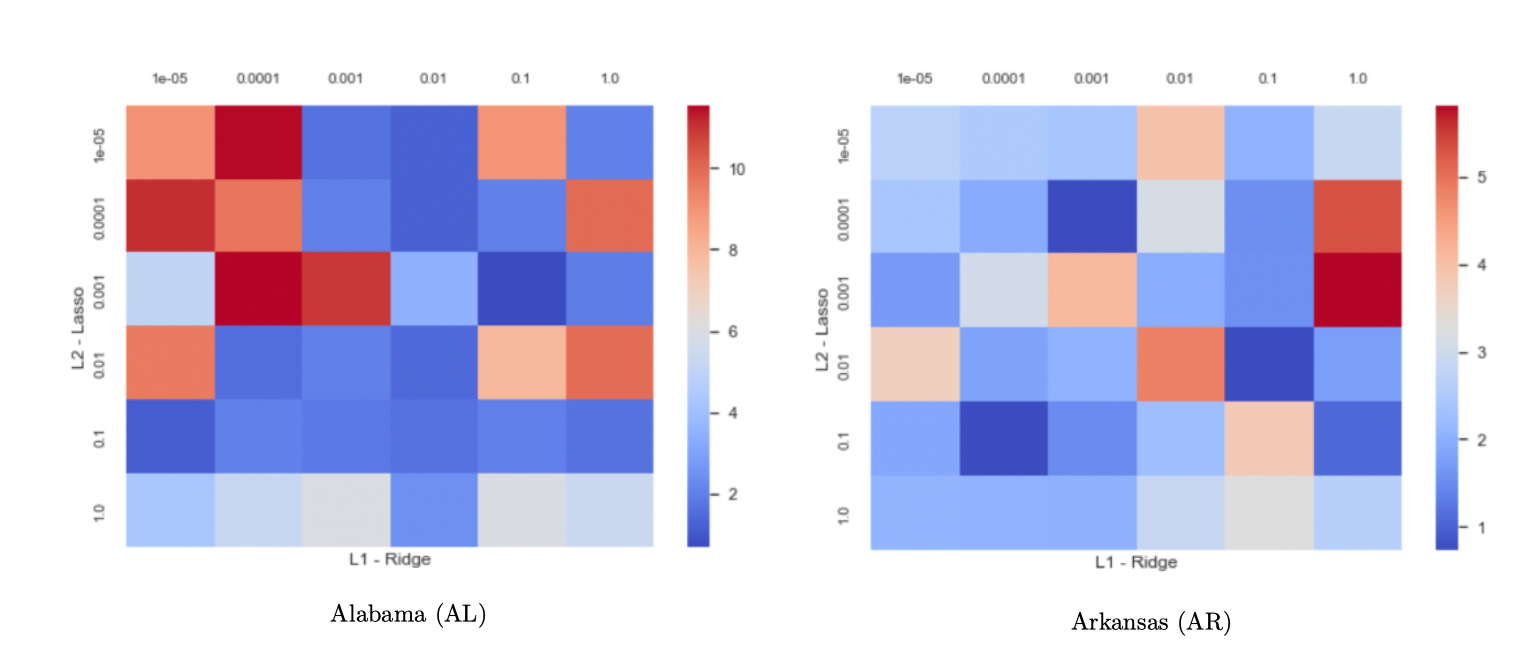}
    \includegraphics[width=16cm]{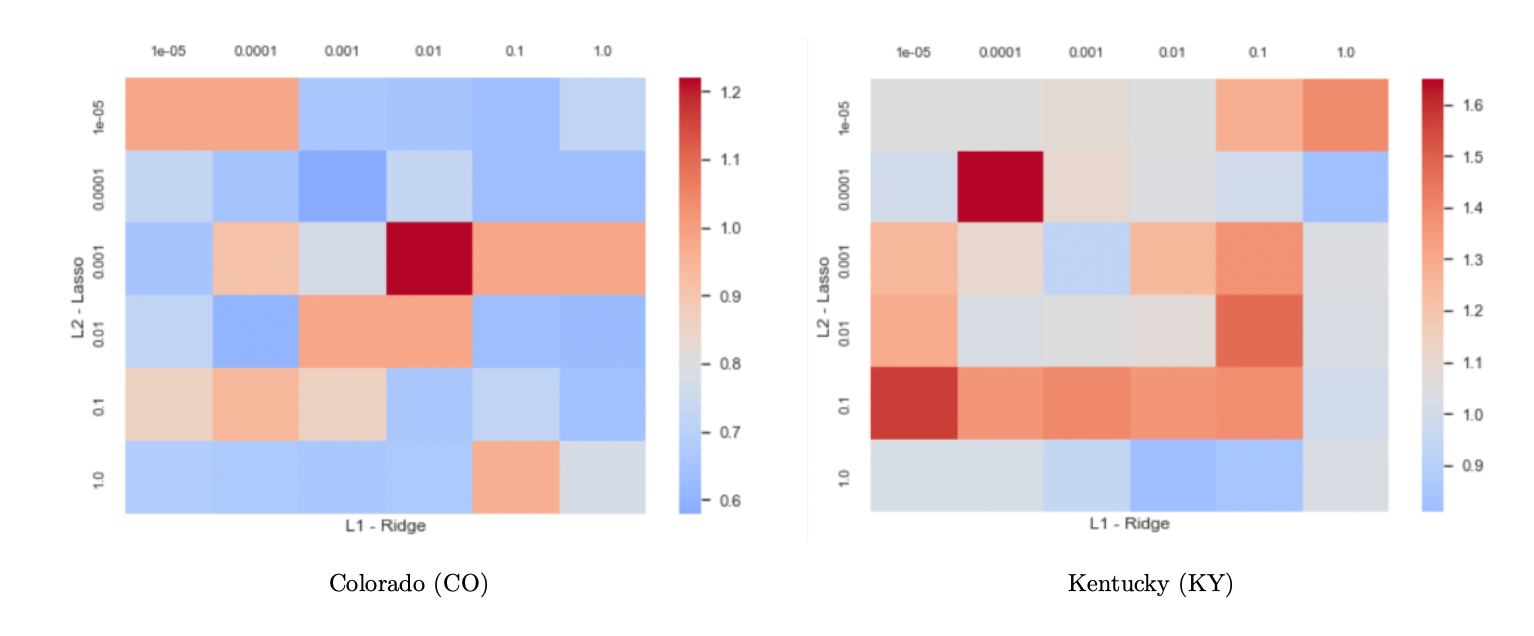}
    \includegraphics[width=16cm]{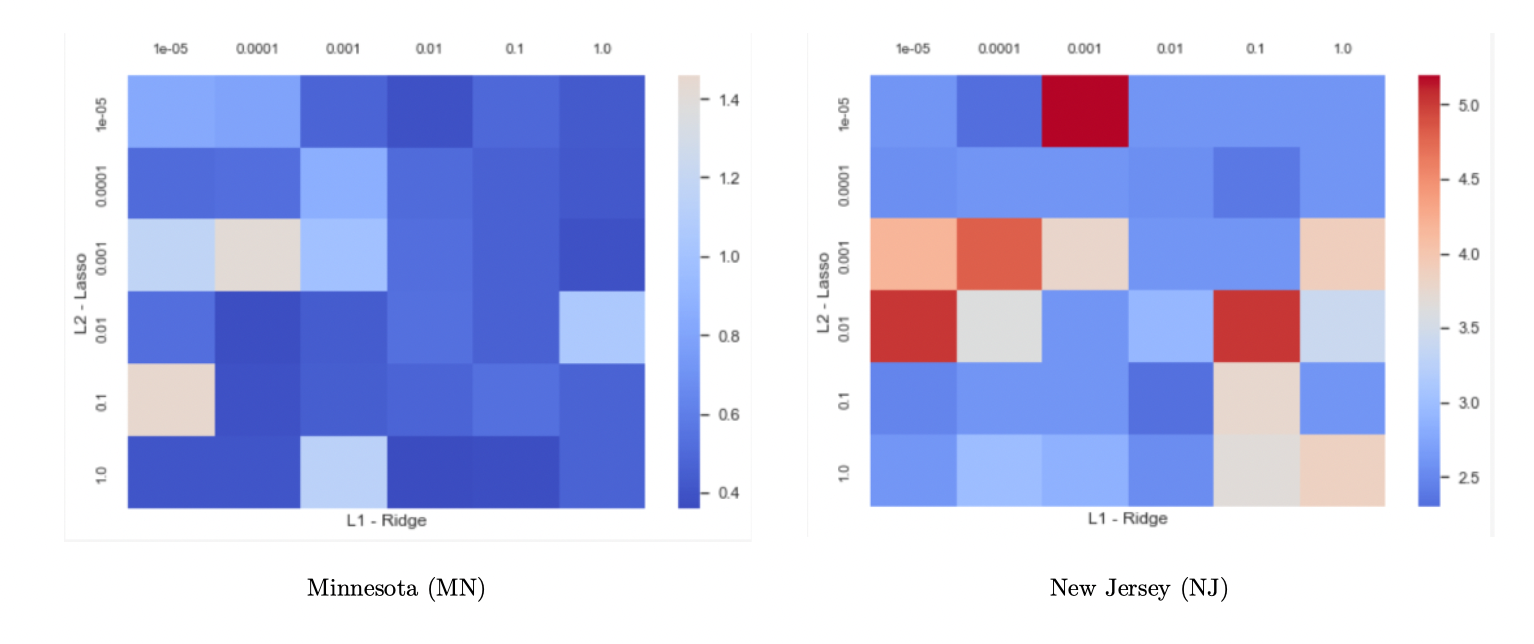}
\end{center}

We notice that the balance is very blurred and it is difficult to draw a clear conclusion. However, if we analyze precisely we notice that the rows $L_1=0.1$ and $L_1=1$ have most often positive effects and we find there among the smallest differences in fit. On the ridge side, no value of $L_2$ stands out and even seems random 
In the notebook, after even more detailed fit tests, we choose the setting: $L_1=0.6$ and $L_2=0.1$.

\subsection{p-values}

In the S7 project we had explained the p-values proposed in Abadie et al (2003), but due to lack of time they had not been calculated. As a reminder, the p-values were constructed in the following way.
\subsubsection{Reminder definition}
To evaluate the quality of the synthetic inference of a processed state (indexed by 1 in the following), we define $0 \leq t_1 \leq t_2 \leq T$ and $j=\{1,... ,J+1\}$,$$R_j(t_1,t_2)=\left(\frac{1}{t_2 -t_1 +1}\sum_{t=t_1}^{t_2}(Y_{jt}-\hat{Y}_{jt}^{N})^2\right)^{1/2}$$ where the terms indexed from 2 to $J+1$ are the quantities specific to each control state.
\\ \\ \indent $R_j(t_1 ,t_2)$ is the RMSE (Root Mean Squared Error) of the synthetic curve compared to the real curve for the state $j$ considered as the treated state, during the period $t_1 , ..., t_2$. In the case where $j=1$, the state is truly treated but for $j>2$, the treated state is a placebo since it is from the control group. We can then define $$r_j=\frac{R_j(T_0+1,T)}{R_j(1,T_0)}$$
$r_j$ measures the quality of the fit of the synthetic control of the entity $j$ over the post-intervention period compared to the quality of that during the pre-intervention period.
\\ \\
We then define the $p$-value of the inference: $$p=\frac{1}{J+1}\sum_{j=1}^{J+1}\bold{1}_{\mathbb{R}+}(r_j - r_1)$$

For the control states (here also called placebo), it is desirable to have a fit error during the pre-intervention period similar to that during the post-intervention period, which translates into $r_j$ close to 1. On the contrary for the treated state, we want the fit error on the post treatment period to be greater than or equal to the error on the pre treatment period. The synthetic curve is therefore all the more relevant if the p-value is close to 0. 
However, the interest of the p-value becomes limited if the synthetic curve of the treated state follows the real curve very closely on the post-treatment period because this implies an $r_1$ close to 1.

\subsubsection{p-values of the model}

\indent In the table below summarizing the value of p-values for different fit times, the primacy of the 10-day fit period is reconfirmed. Moreover, the quality of the estimates by state seems to be stable: California, Kentucky, Maryland, Michigan, Minnesota, New York, Ohio and West Virginia have a p-value below 0.25. Other states with higher p-values, up to 0.90 for Maine, have less reliable results.

\begin{center}
    \includegraphics[width=15cm]{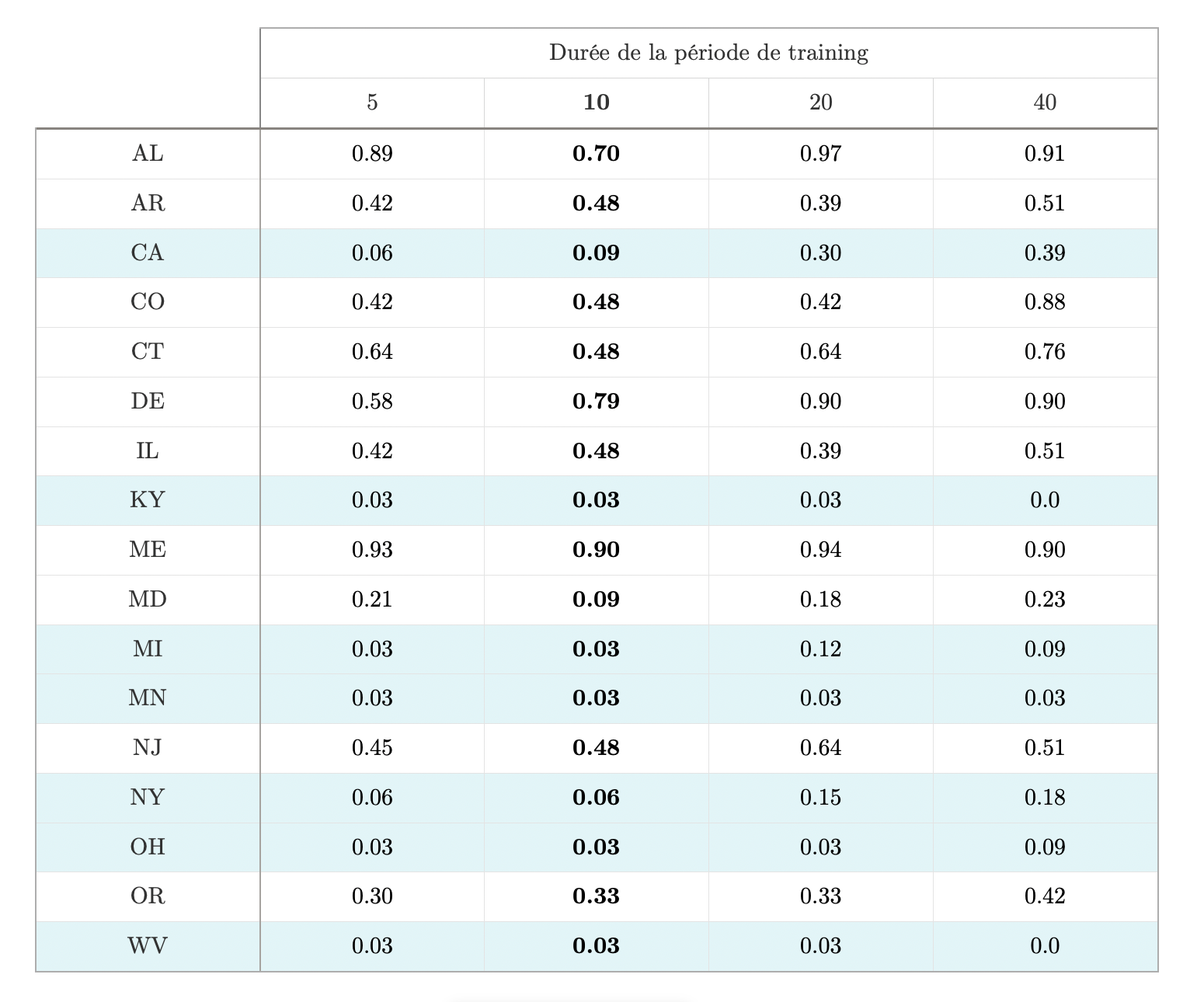}
    \textit{Figure 2: p-values (Note: the start date of the incentive $t_0$ chosen for the placebo control states is Illinois (5/26/2021))}
\end{center}

\subsection{Return on the results}

These improvements and additional information allow us to produce more reliable results that we will analyze.
\\ \\ 
\indent We will limit our analysis to the whole population instead of a particular age group.  The results for the age groups are first of all of much better quality than in the previous semester. The same conclusions can be drawn in most cases for each age group, so the interest of the analysis is limited. All of the following results as well as those for the age groups are available on the interactive dashboard. 
\\ \\
\indent To be able to report the most interesting conclusions possible, we selected for each state the fit time that delivers the best fit over the pre-intervention period. This approach is acceptable because the results are overall very stable per change of fit time, a clear improvement over the S7 results.
\\ \\

\subsubsection{Results table}
The results are summarized in the table below:
\\
\begin{longtable}{|m{1.8cm}||c|m{5cm}|m{5cm}|c|}
\hline
    Processed state & p-value & Analysis & Curve & t\_fit\\
    \hline \hline
   Alabama (AL) & 0.91 & The fit is not stable for the different fit times, so the curve is not interpretable. Indeed, the setting t\_fit=40 is the only one which allows a good pre-intervention fit. The p-value is a good indicator of this lack of reliability. & \includegraphics[width=5cm]{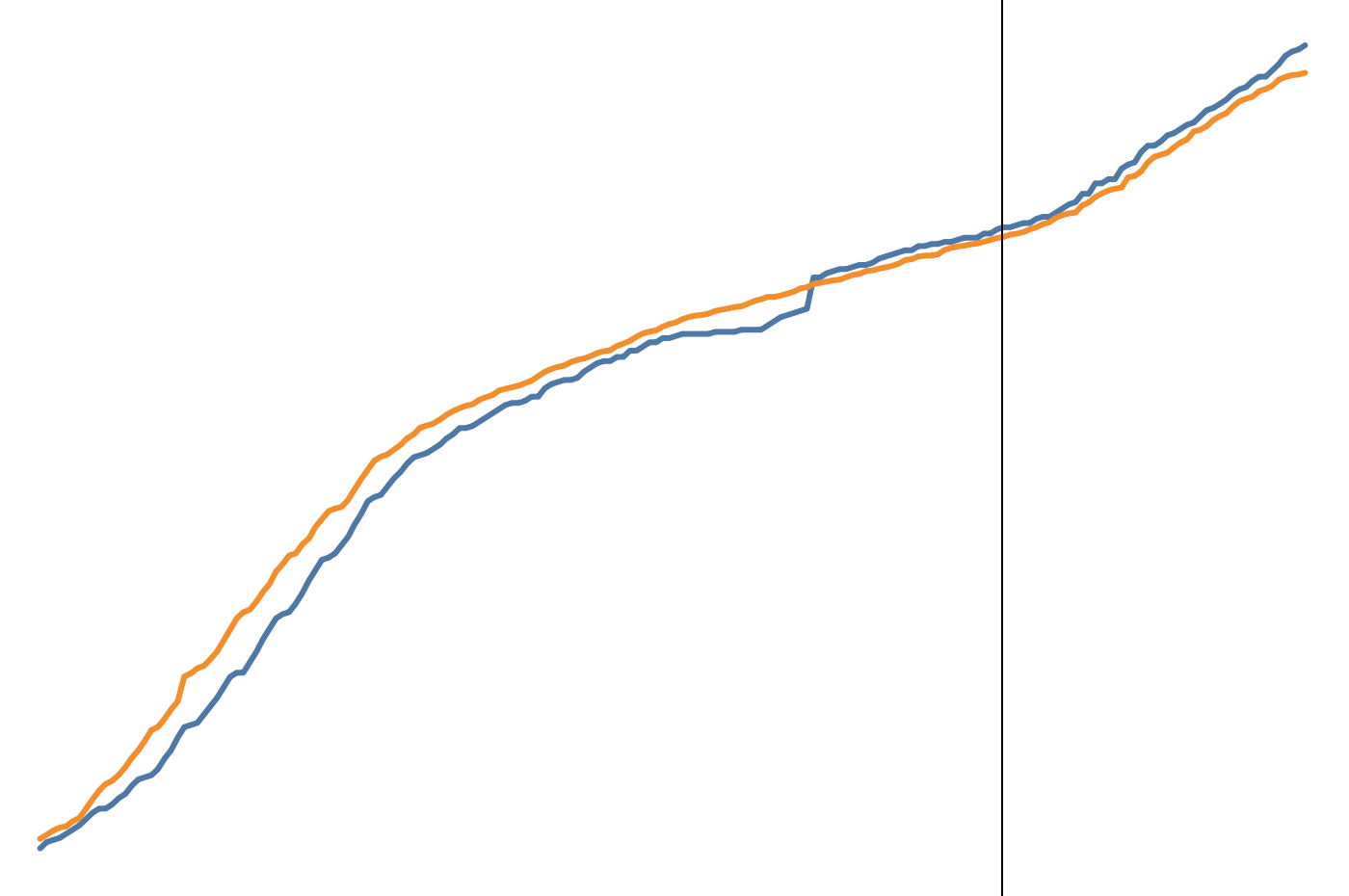} & 40\\
   \hline
   Arkansas (AR) & 0.48 & There is a bad fit. We can't conclude anything. & \includegraphics[width=5cm]{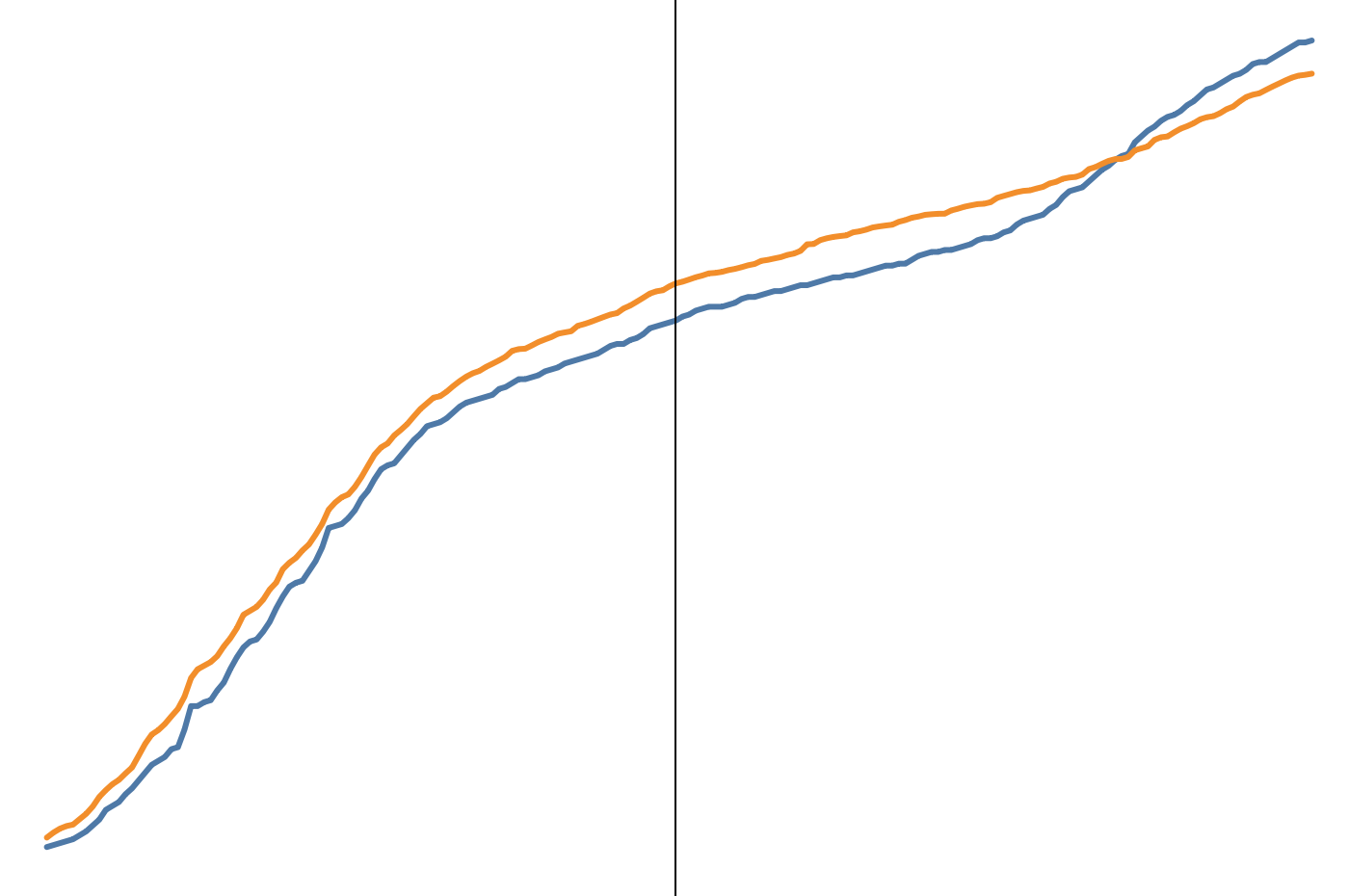} & 10\\
   \hline
   California (CA) & 0.09 & The p-value indicates that the result is significant. There is a slight fit problem around the months of February and March but this is negligible compared to the good fit just before the intervention. In view of the post-treatment period, we conclude that the incentive had no effect.  & \includegraphics[width=5cm]{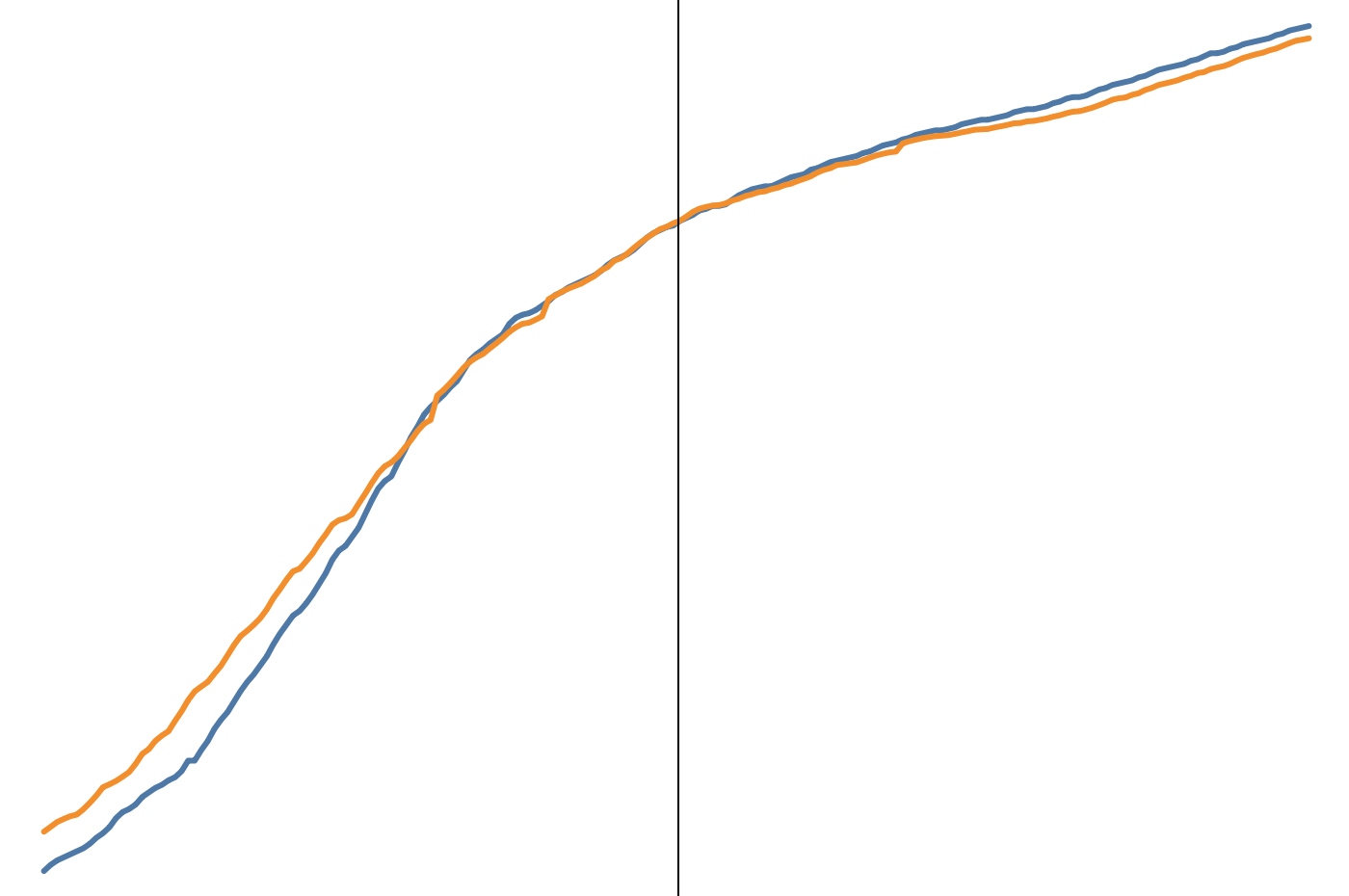} & 10\\
   \hline
   Colorado (CO) & 0.42 & The synthetic is interesting because it seems to indicate a lack of effect of the incentive. However the relatively high p-value makes the result moderately reliable. Nevertheless, we can assume that the p-value is so because of the proximity of $r_1$ to 1 (the synthetic and the real are almost identical over the whole period of the study), which could eventually legitimize our result.   & \includegraphics[width=5cm]{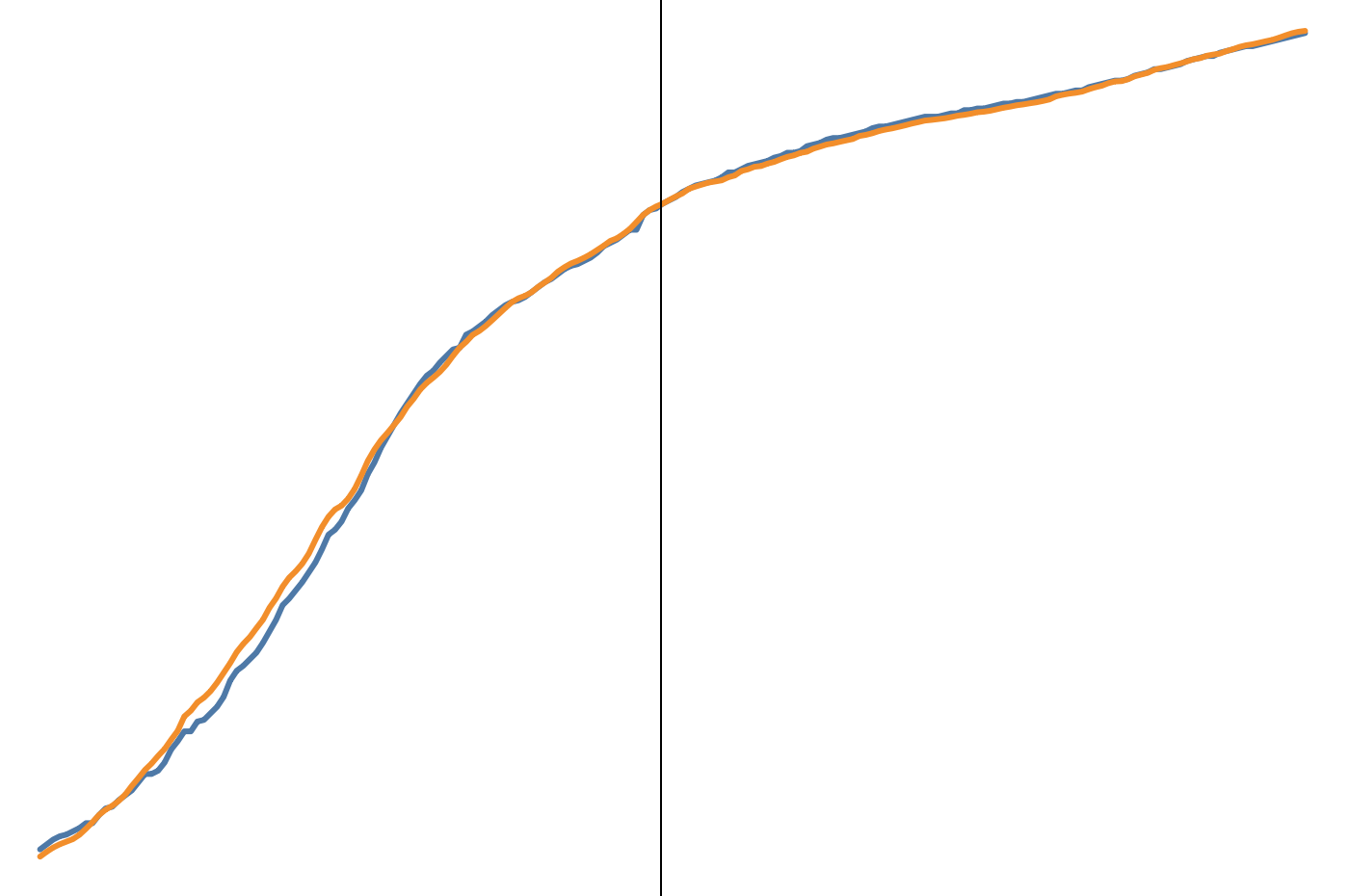} & 20\\
   \hline
   Connecticut (CT) & 0.76 & There is a bad fit. We can't conclude anything. & \includegraphics[width=5cm]{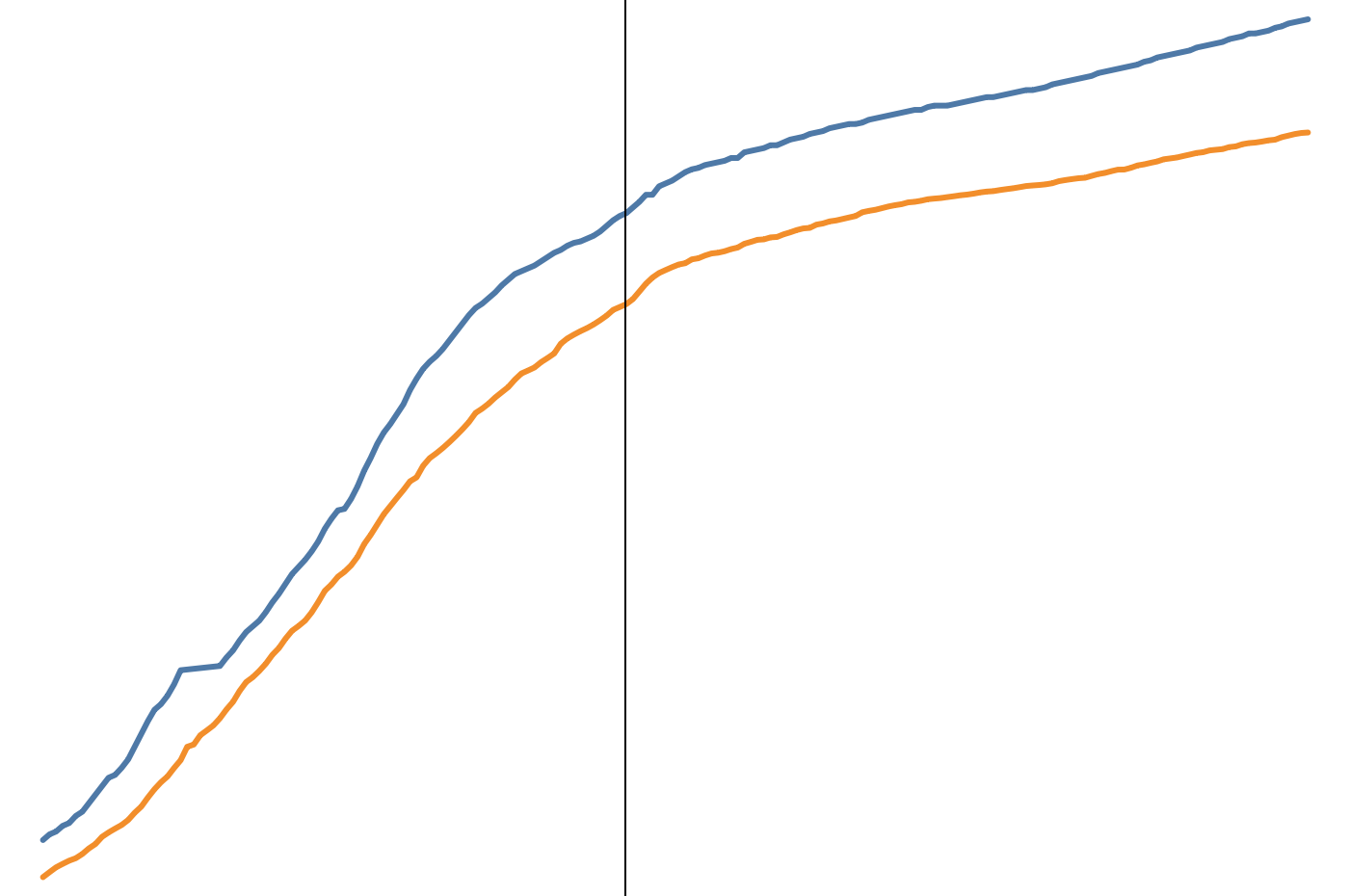} & 40\\
   \hline
   Delaware (DE) & 0.90 & The synthetic is interesting because it seems to indicate a lack of effect from the incentive but the high p-value indicates that the study is insignificant.  & \includegraphics[width=5cm]{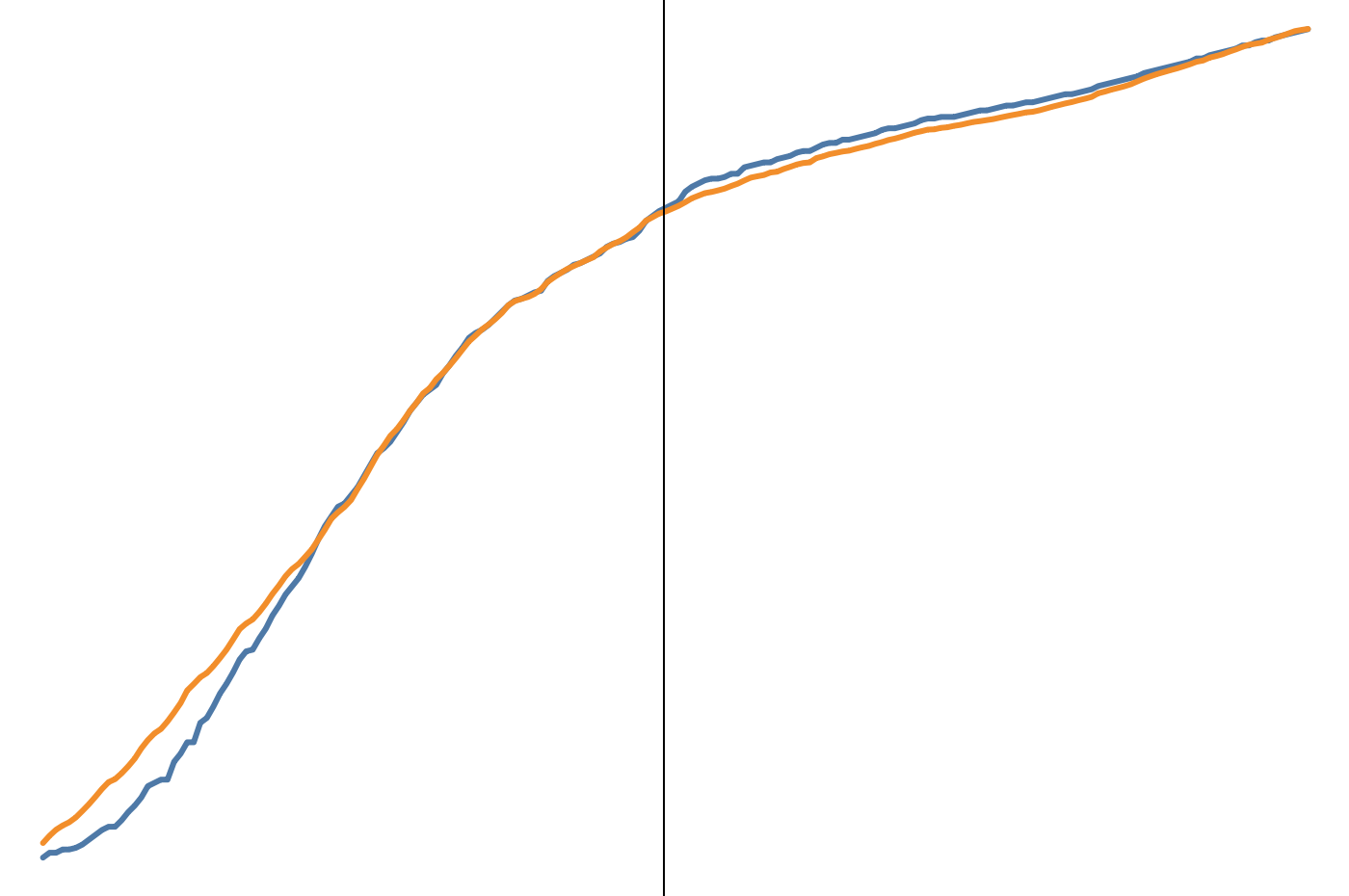} & 40\\
   \hline
   Illinois (IL) & 0.48 & There is a bad fit. We can't conclude anything. & \includegraphics[width=5cm]{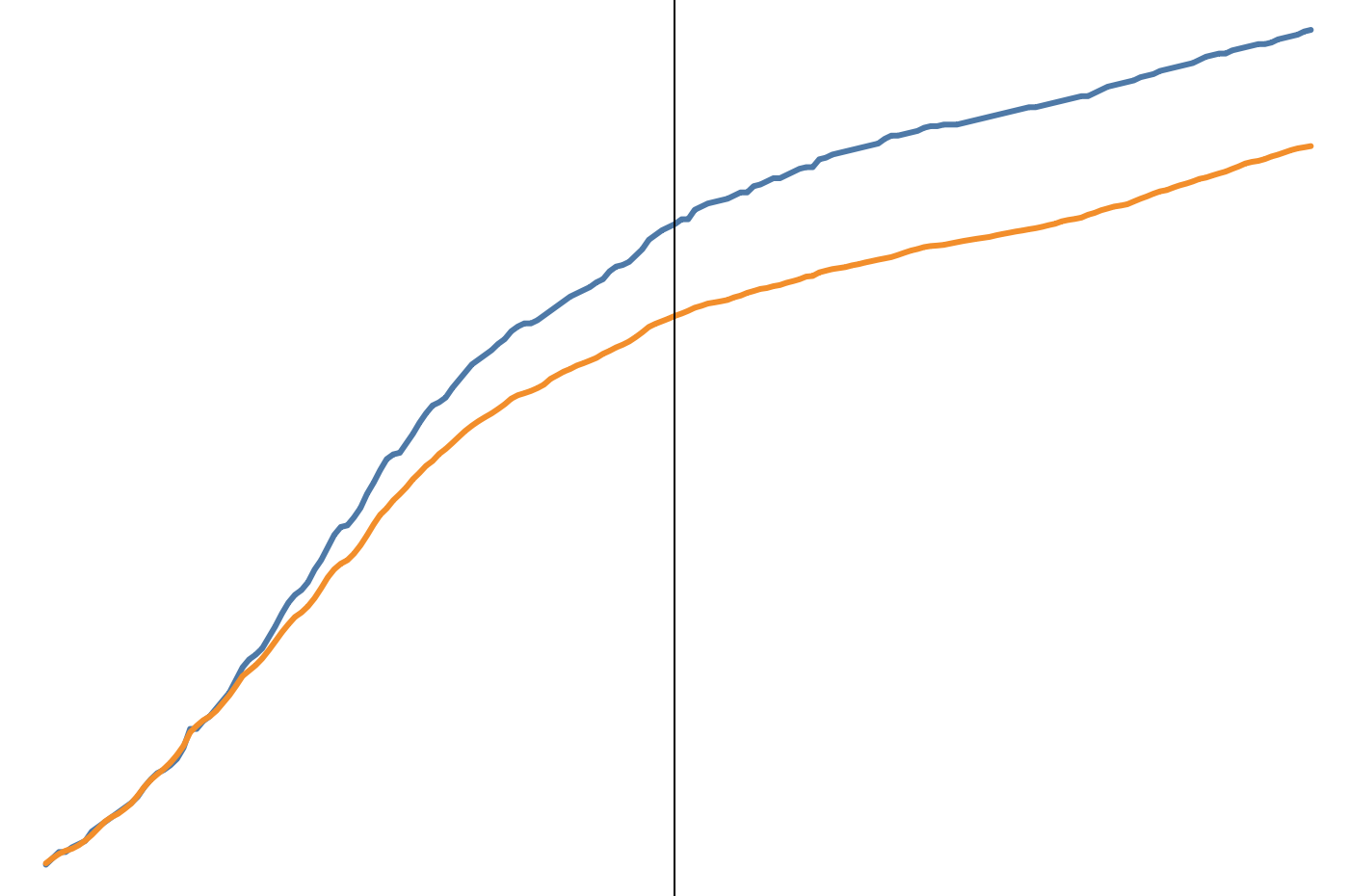} & 10\\
   \hline
   Kentucky (KY) & 0.03 & In view of the post-treatment period and the low p-value, it is concluded that the incentive had no effect. & \includegraphics[width=5cm]{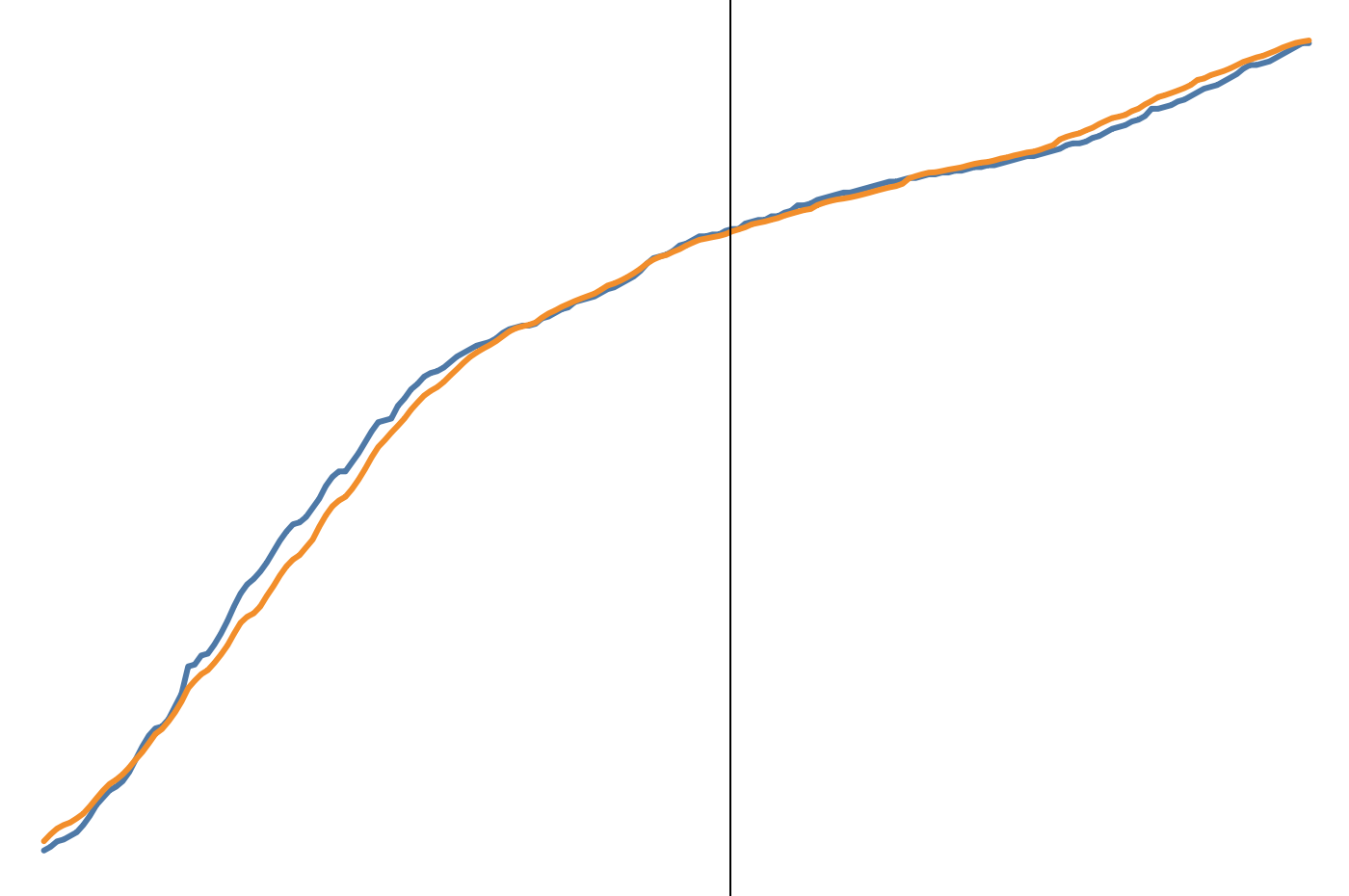} & 10\\
   \hline
   Maine (ME) & 0.90 & Maine's high p-value is felt in the variable fit quality of the synthetic. However, the measure does not really seem to have an effect, although this result is weakly supported by the p-value.& \includegraphics[width=5cm]{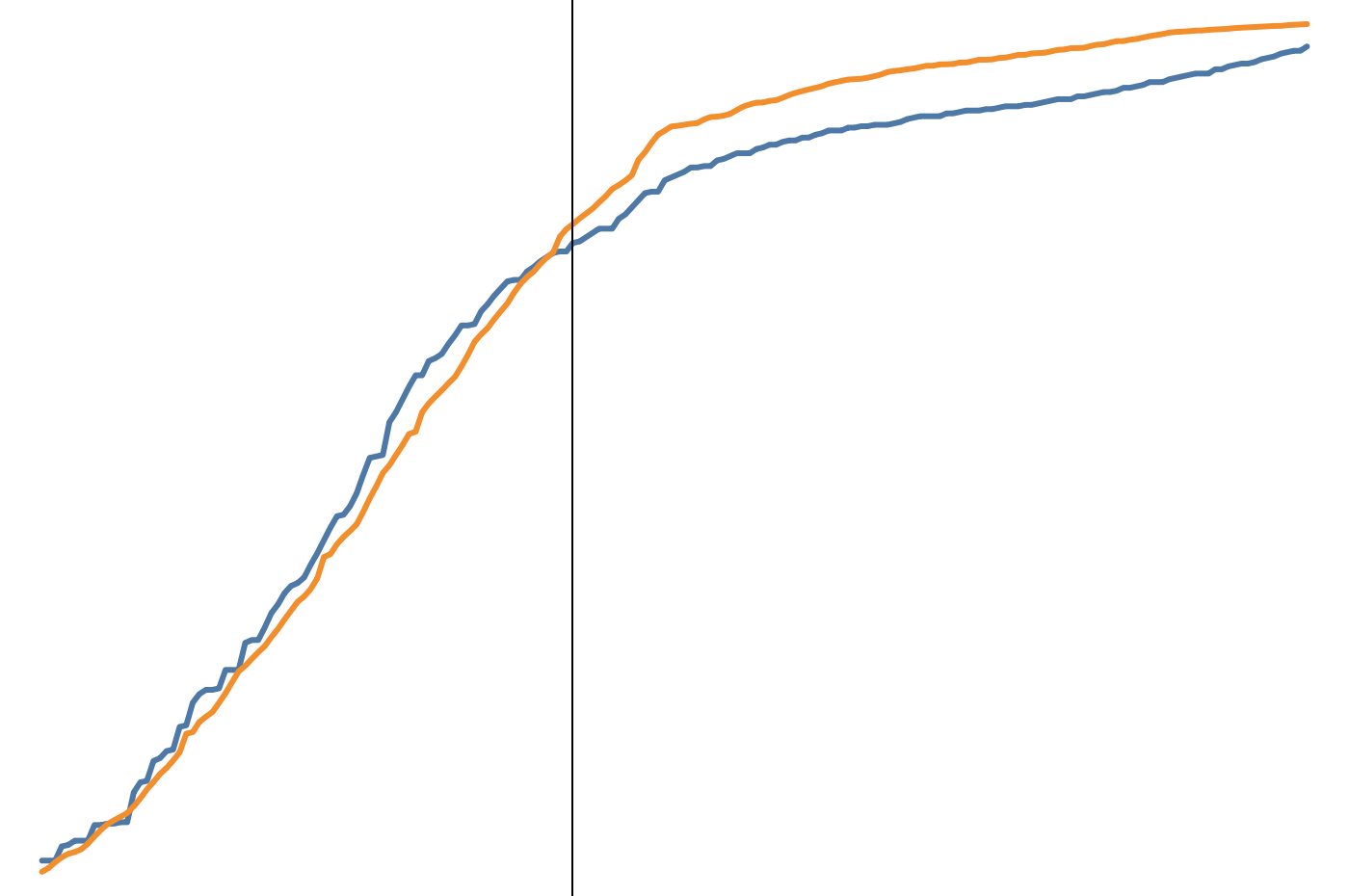} & 40\\
   \hline
   Maryland (MD) & 0.09 & The fit is good. No effect of the incentive is noticed immediately, but rather after about 12 days. This does not, however, correspond to any additional incentive being implemented. The low p-value supports the legitimacy of these results. & \includegraphics[width=5cm]{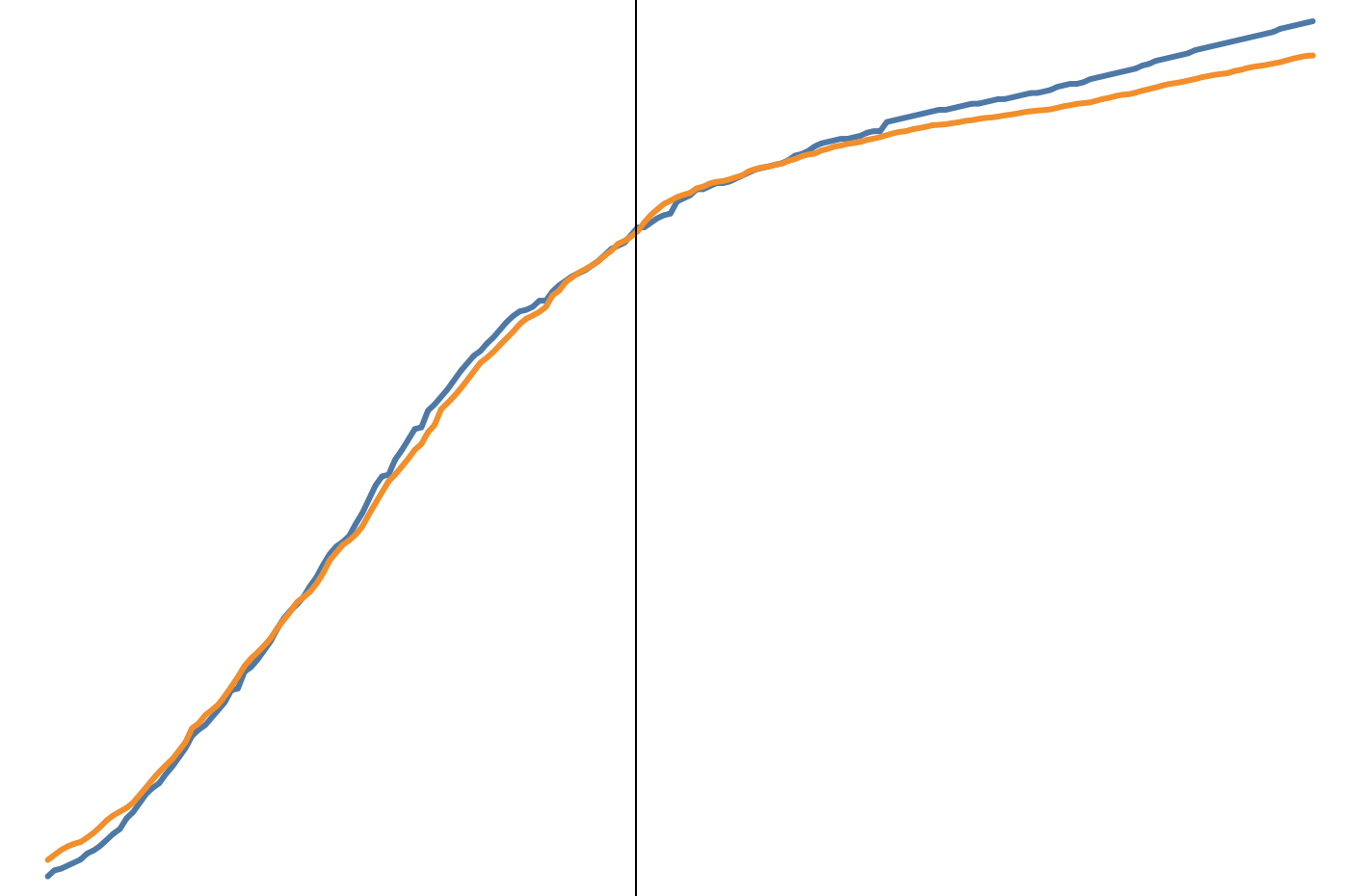} & 10\\
   \hline
   Michigan (MI) & 0.03 & In view of the post-treatment period and the low p-value, we conclude that the incentive had no effect. We even notice that the rate of vaccination slowed down more than in similar states.& \includegraphics[width=5cm]{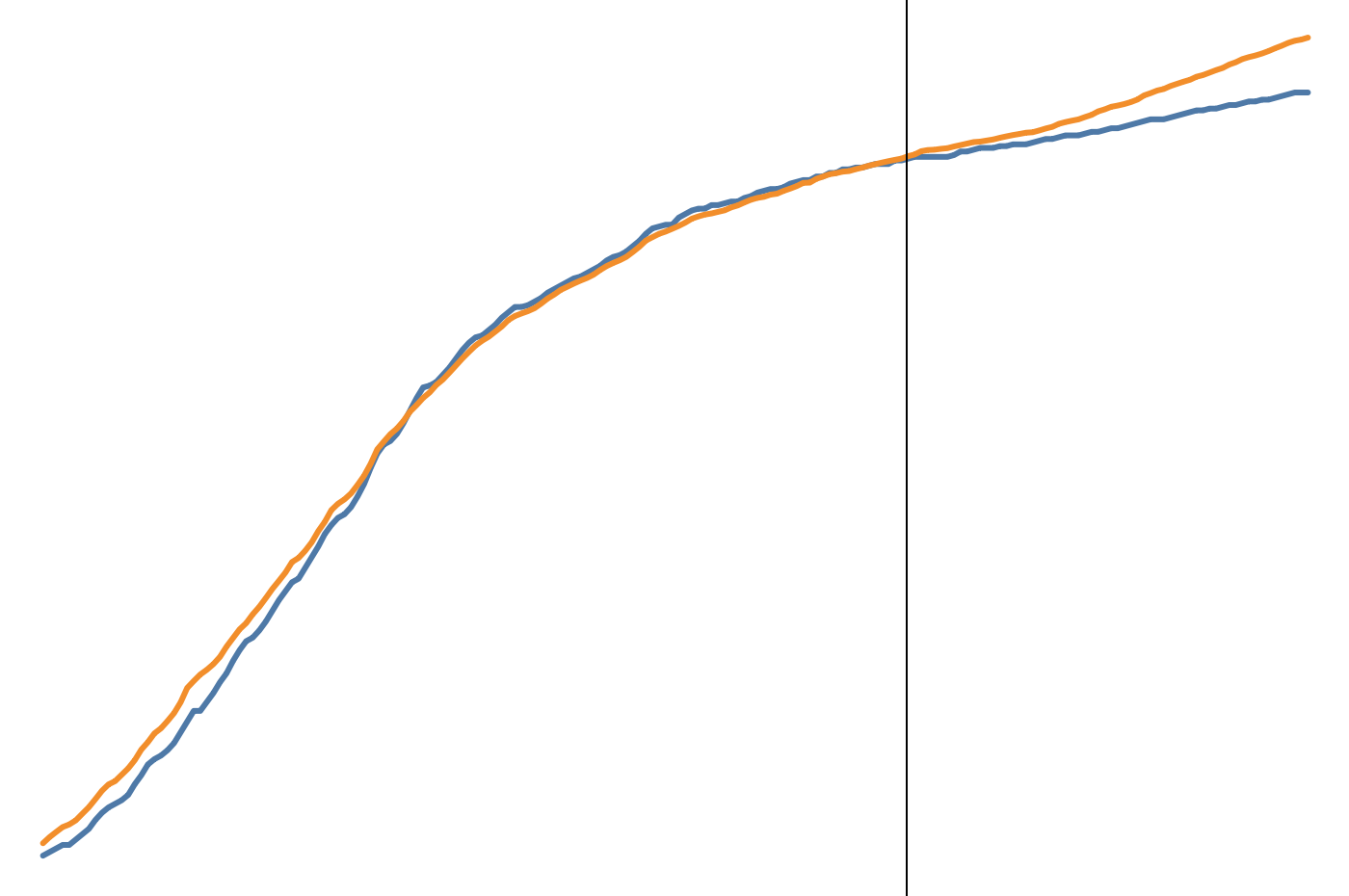} & 10\\
   \hline
   Minnesota (MN) & 0.03 & In view of the post-treatment period and the low p-value, we conclude that the incentive had no effect. Indeed, the difference after $t_0$ can be considered negligible. & \includegraphics[width=5cm]{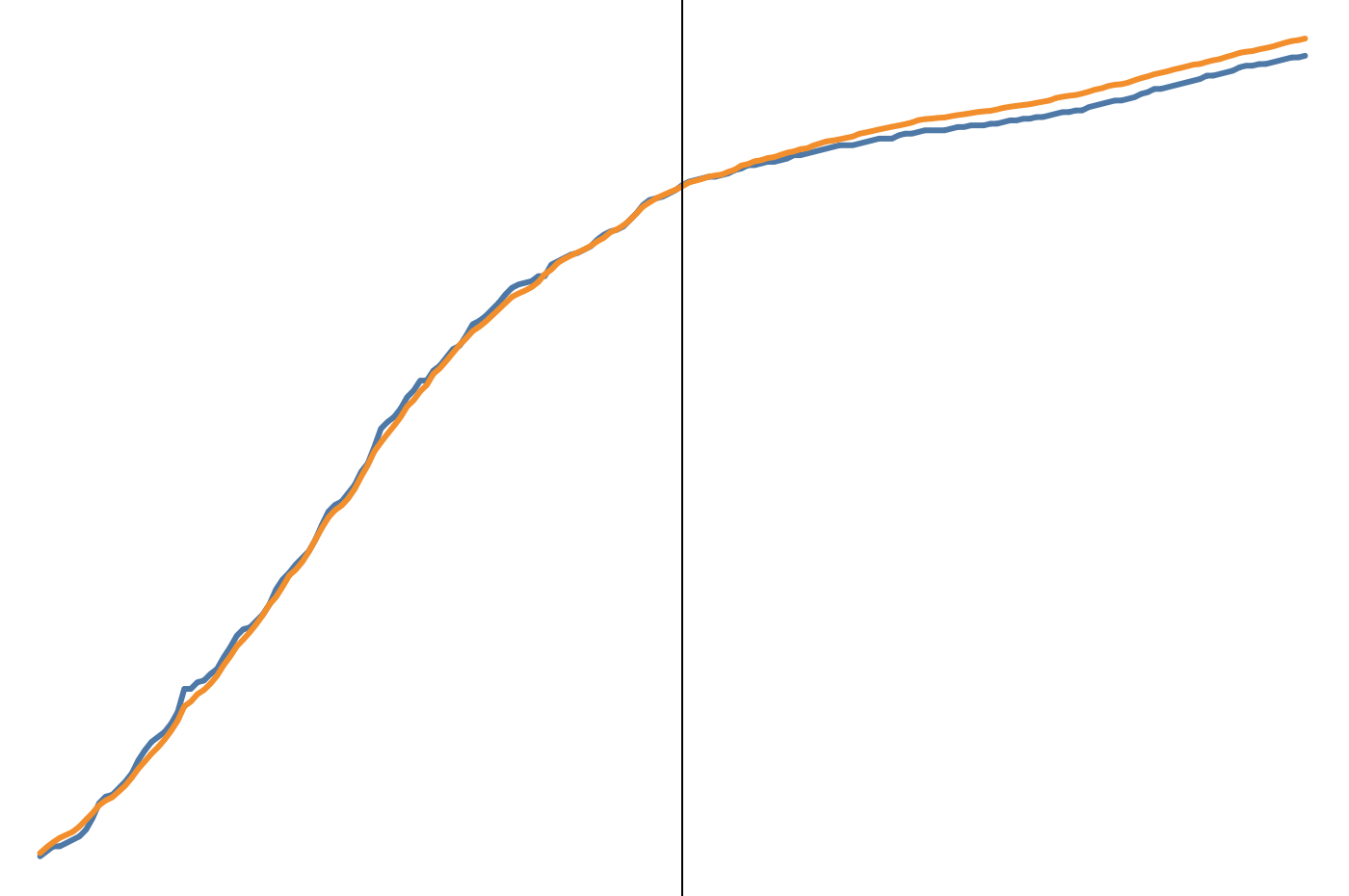} & 10\\
   \hline
   New Jersey (NJ) & 0.64 & There is a bad fit. We can't conclude anything. & \includegraphics[width=5cm]{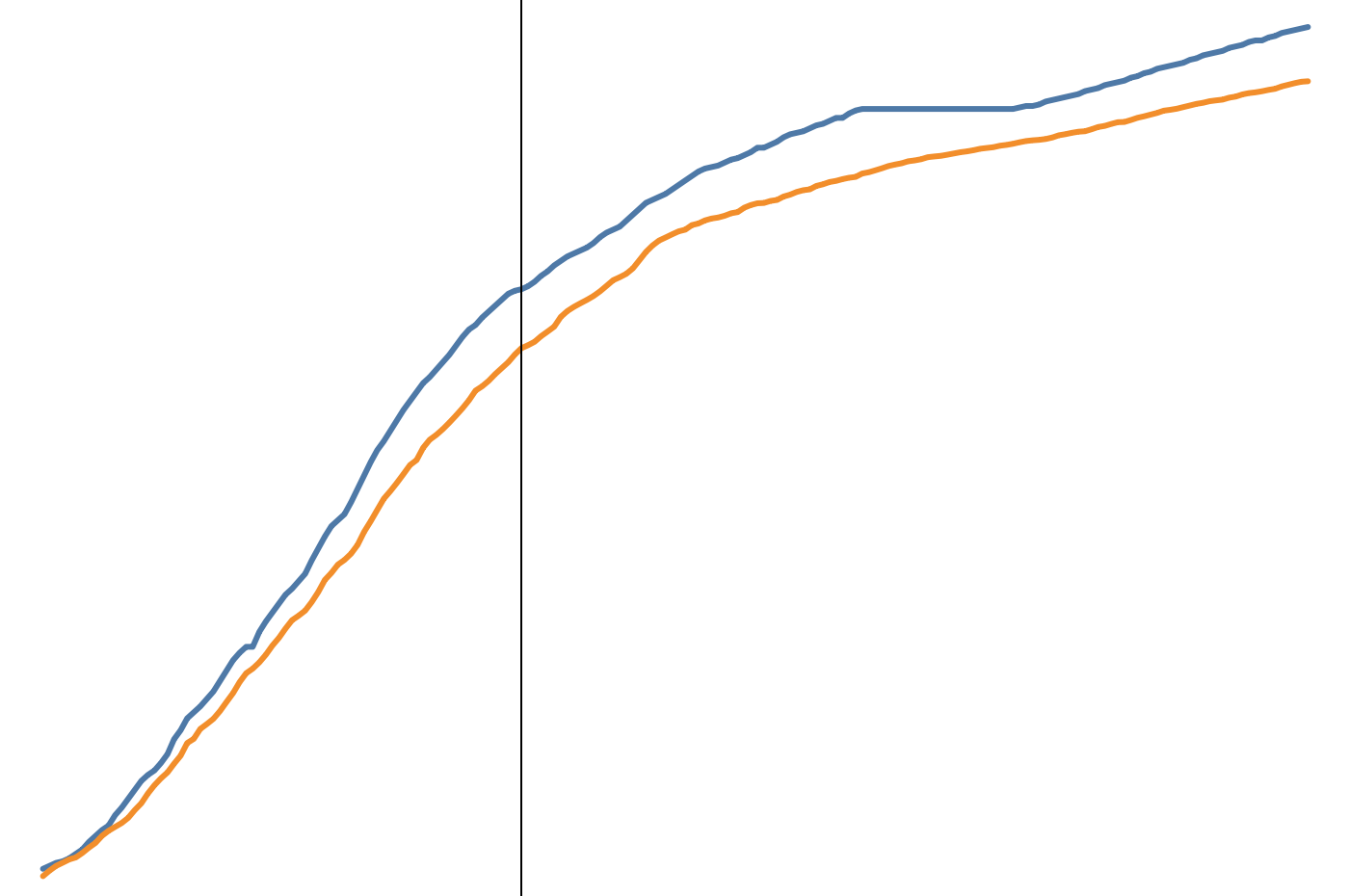} & 20\\
   \hline
   New York (NY) & 0.18 & The fit is good. No effect of the incentive is seen immediately, but rather about a fortnight later, which corresponds to the implementation of several incentive lotteries. The low p-value supports these results.  & \includegraphics[width=5cm]{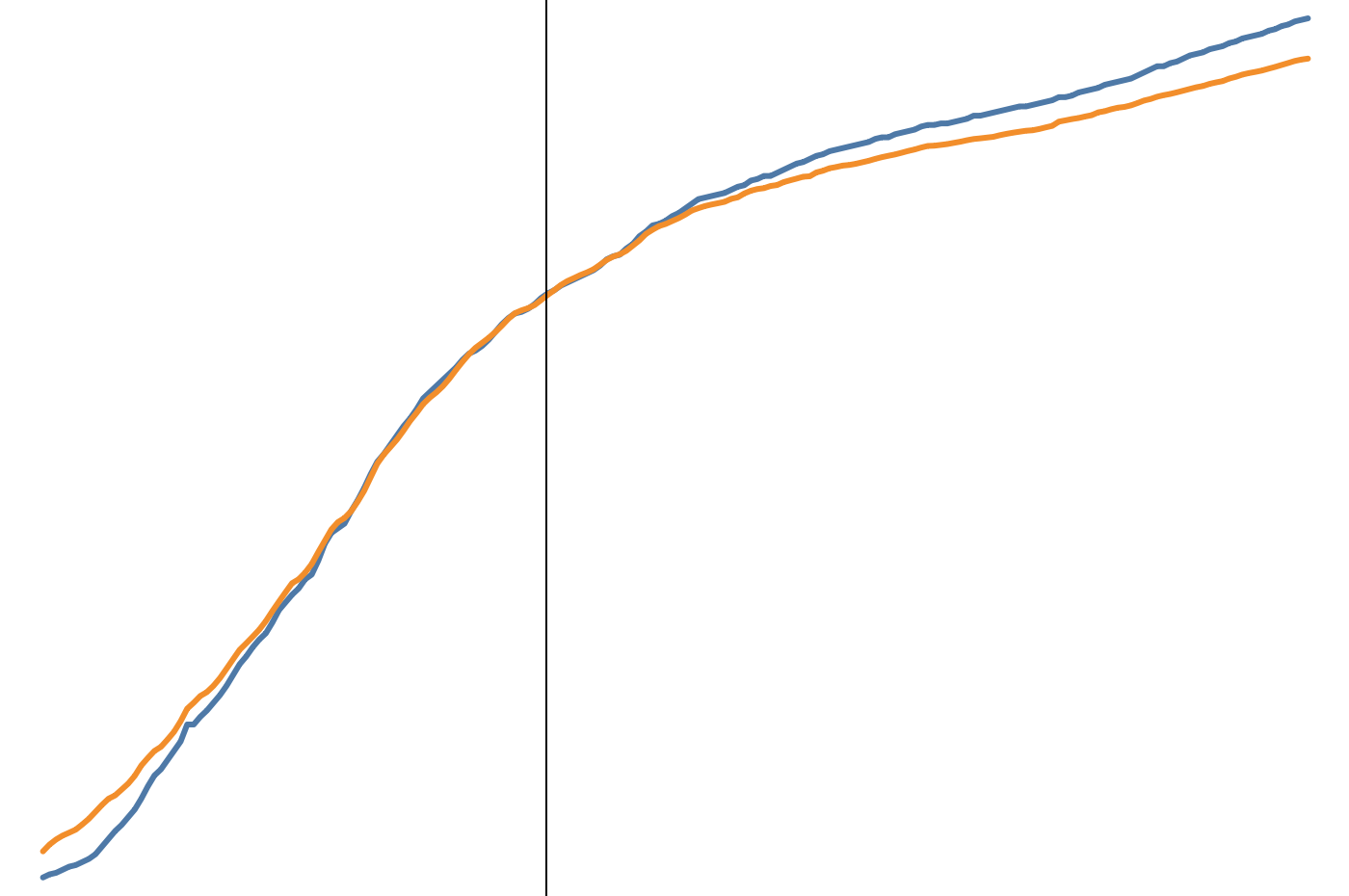} & 40\\
   \hline
   Ohio (OH) & 0.03 & Based on the post-treatment period and the low p-value, it is concluded that the incentive had no effect. & \includegraphics[width=5cm]{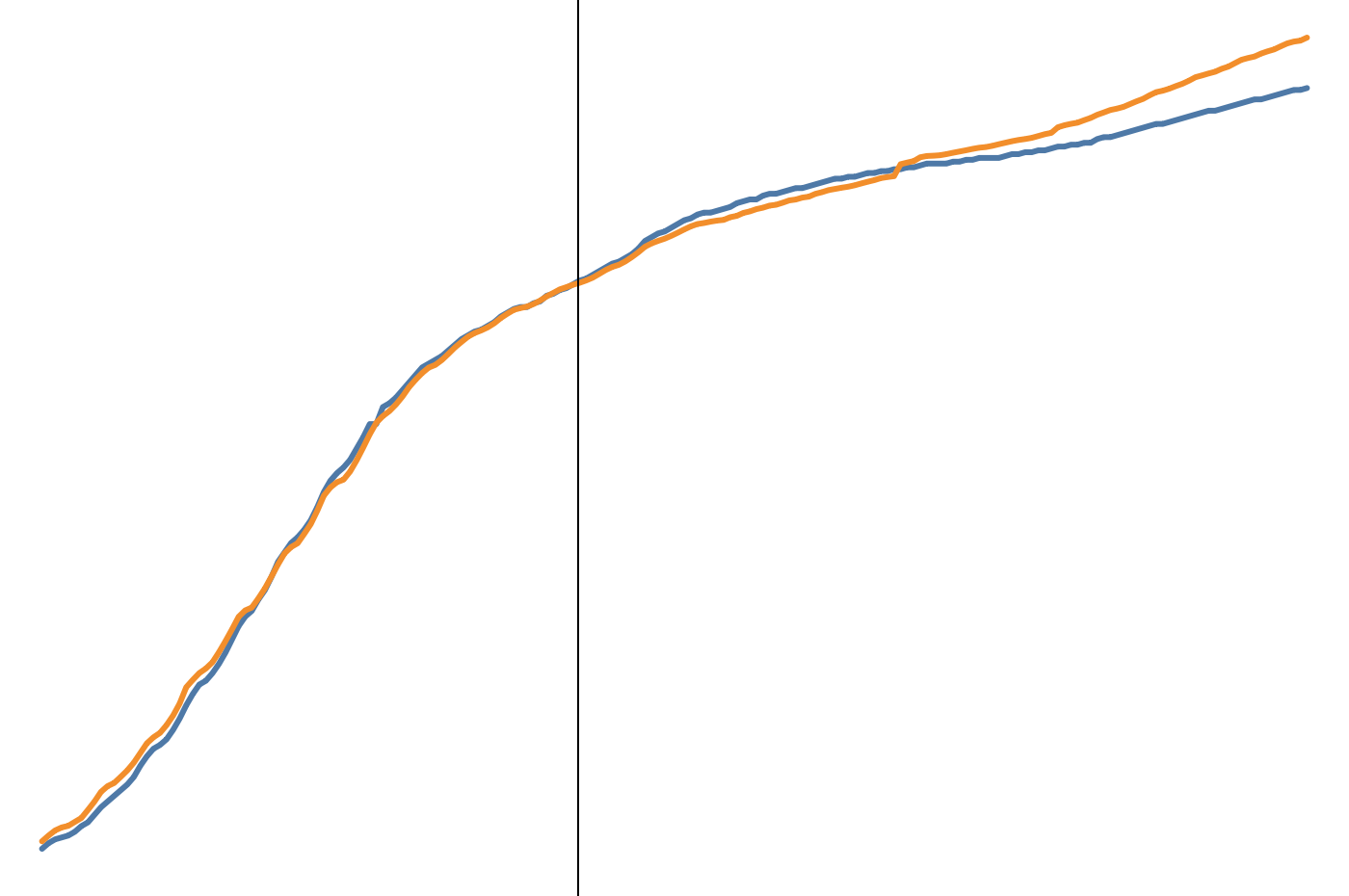} & 10\\
   \hline
   Oregon (OR) & 0.42 & There is a bad fit. We can't conclude anything. & \includegraphics[width=5cm]{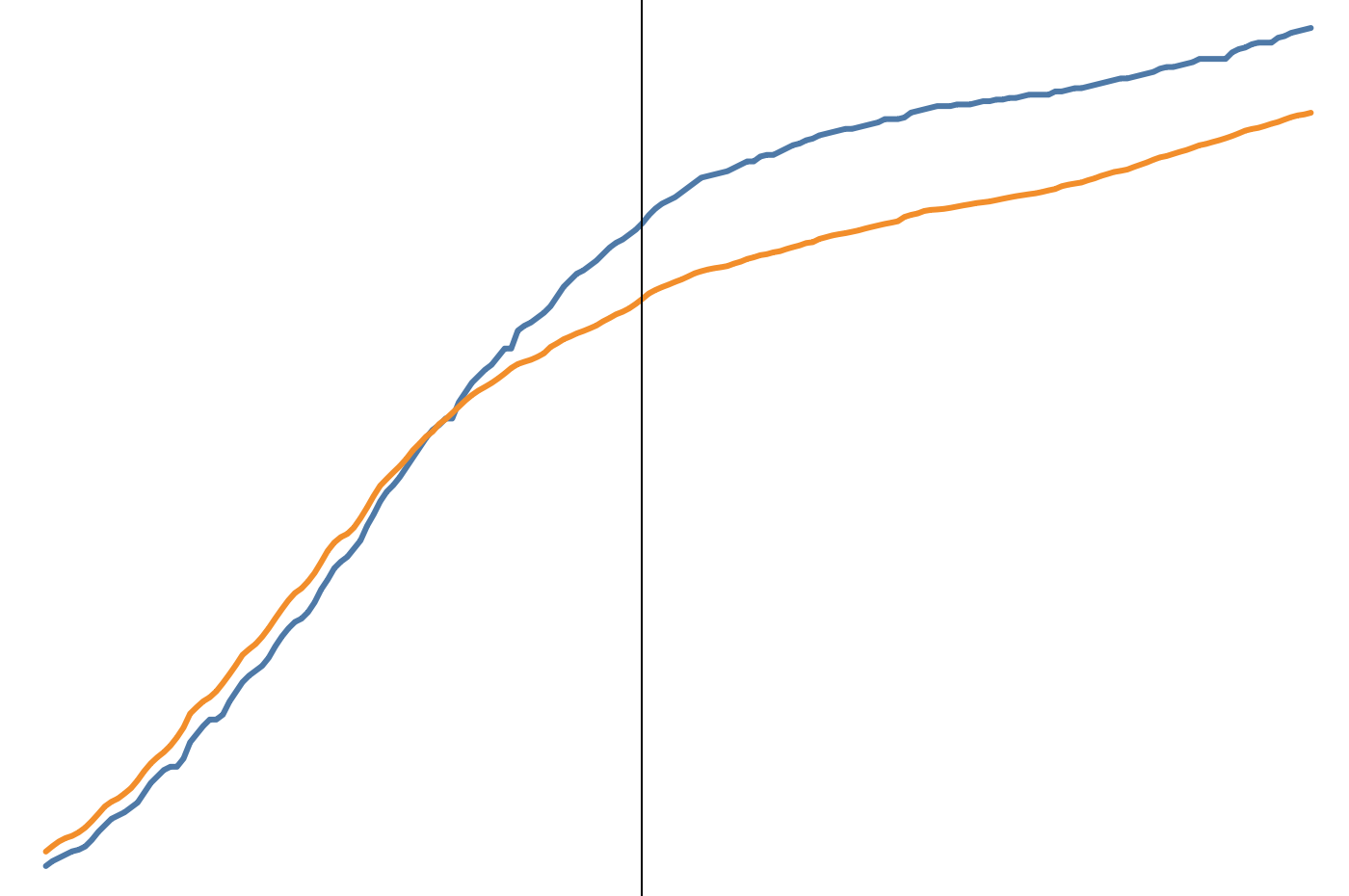} & 40\\
   \hline
   West Virginia (WV) & 0.0 & The fit is good in the pre-intervention period. It is therefore interesting to note that the incentive had no significant effect. Despite this, the zero p-value is exclusively due to the large divergences towards the end of the pre-intervention period. The low quality of the data for this condition forces us to view these results with caution, without invalidating them.& \includegraphics[width=5cm]{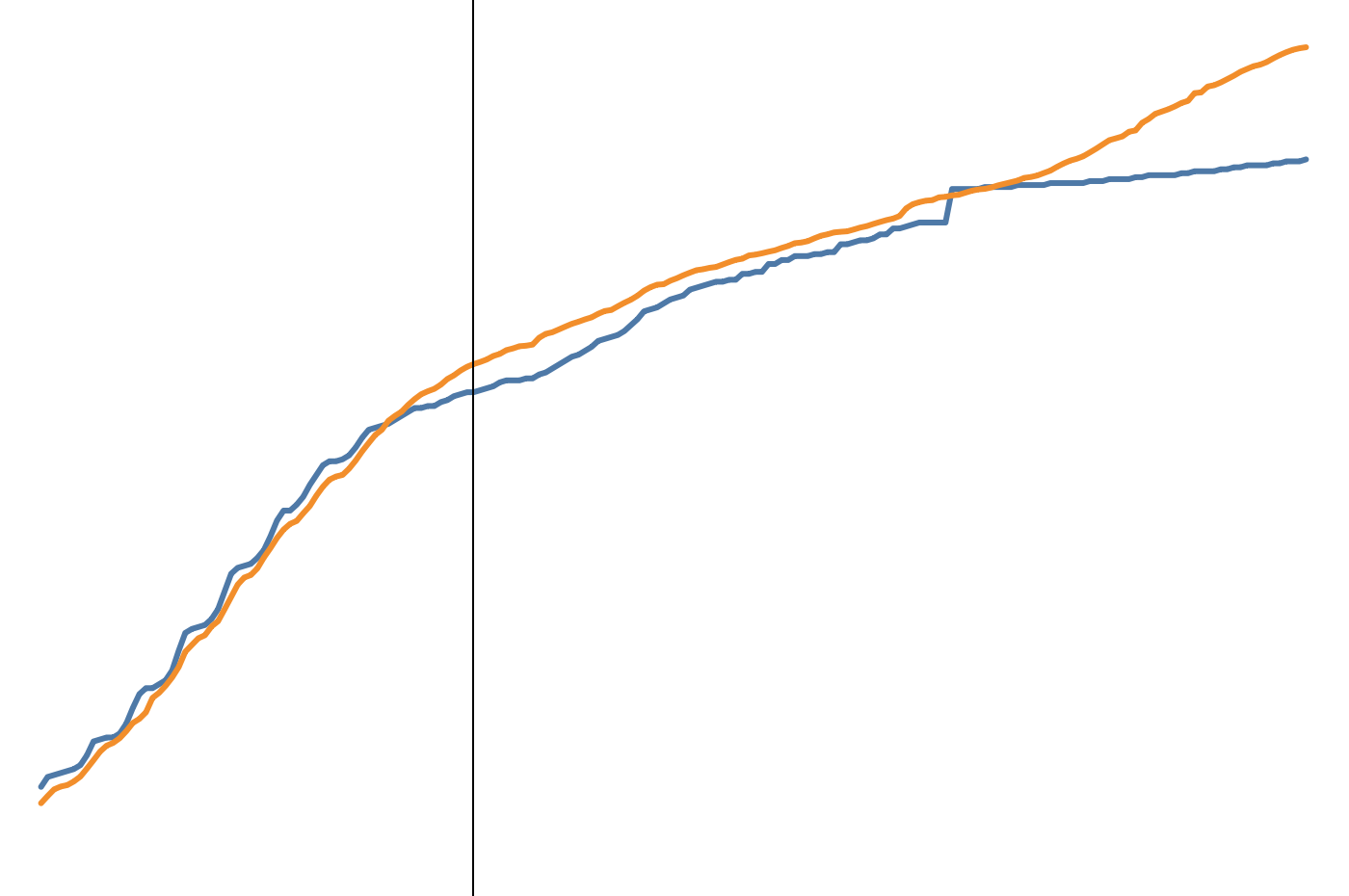} & 40\\
   
   \hline
   

 \end{longtable}
  
\subsubsection{Conclusion on the effectiveness of statewide incentives}

With these improved results, nearly half of the states have clear conclusions about the effectiveness of incentives. With the exception of New York and Maryland (which now require more detailed study), regardless of the type of incentive implemented (gift cards, lotteries, etc.), there is no acceleration in the daily number of citizens getting their first dose of Covid-19 vaccine. Thus, incentives do not appear to have had a significant effect on vaccination in the states overall. \\ \indent One could possibly assume that these measures have kept vaccinations at the same levels as other states without incentives, but this remains speculative and is difficult to assess.

\newpage

\section{Socio-demographic study of the populations}

\subsection{Motivation}

\indent While we have been able to estimate and judge the relevance of the different incentives in the United States to increase the overall vaccination rate of the country, we must now try to understand more precisely which are the socio-demographic factors that influence the vaccination of a population. To do this, we used the study [6] in California of the different vaccination dynamics of the Covid-19 according to socio-demographic vulnerability indices. We generalized the study to all American states and counties, trying to understand the different correlations between the vulnerability indices of the populations and their ability to be vaccinated. 

\subsection{Data presentation}

\indent In fact, we have data both on the socio-demographic characteristics of the communities, which can be broken down into several themes, and on the parameters of their vaccination. 

\subsubsection{Community Vulnerability Index (CVI)}

\indent In the context of our study, it is appropriate to present the set of themes addressed in order to measure the vulnerability index of each county in the United States. Indeed, the definition of inequalities and disparities with regard to access to health care or even other privileges is not always easy to characterize and indeed to measure. Thus, based on the task published in the Journal of Immigrant and Minority Health [6], the aim is to group together six different indices that could potentially influence vaccination rates.

\begin{itemize}
    \item[$\bullet$] \textbf{Theme 1 : Socio-economic status}. This takes into account the total population living below the poverty line, those who are unemployed, the wage distribution, and also the proportion with a diploma at the end of high school. 
    \item[$\bullet$] \textbf{Theme 2: Household composition and disability}. This corresponds to the age distribution of households, but also takes into account the presence of people with disabilities and the number of single-parent families.
    \item[$\bullet$] \textbf{Theme 3: Status and language of minorities}. The ability of minorities in populations to speak English "Less than well". 
    \item[$\bullet$] \textbf{Theme 4: Type of housing and transportation}. Lists the types of houses, such as multi-unit structures, mobile homes, but also the absence of vehicles for example.
    \item[$\bullet$] \textbf{Theme 5: Epidemiological factors}. In this part we take into account the distribution of different diseases in the population such as diabetes, obesity or cardiovascular problems. 
    \item[$\bullet$] \textbf{Theme 6: Health System Factors}. For this last theme, the access, capacity and quality of surrounding hospital services are considered.
\end{itemize}

Finally, there is also a global index taking into account all 6 themes.

\subsubsection{Vaccination parameters}

In addition to the socio-demographic characteristics, the model used in the article on Californian counties [6] defines two parameters to determine the shape of their vaccination rate curve. To model them, we represent them by a logistic growth curve since the vaccination rate actually reaches an asymptotic value. Thus, the two parameters of the model should be defined:

\begin{itemize}
    \item[$\bullet$] \textbf{K} : which represents the asymptotic maximum that the vaccination rate can reach
    \item[$\bullet$] \textbf{$\nu$} : which corresponds to the vaccination velocity, the speed at which the rate reaches its steady state.
\end{itemize}

Thus, the equation for the vaccination rate in a county at time t can be given :
$$
p_t = \frac{Kp_0e^{\nu t}}{K + p_0(e^{\nu t} - 1)}
$$
However, in order to get an idea of the distribution of the two parameters across the United States, it is worth looking at their distribution by state.
\begin{center}
    \includegraphics[width=12cm]{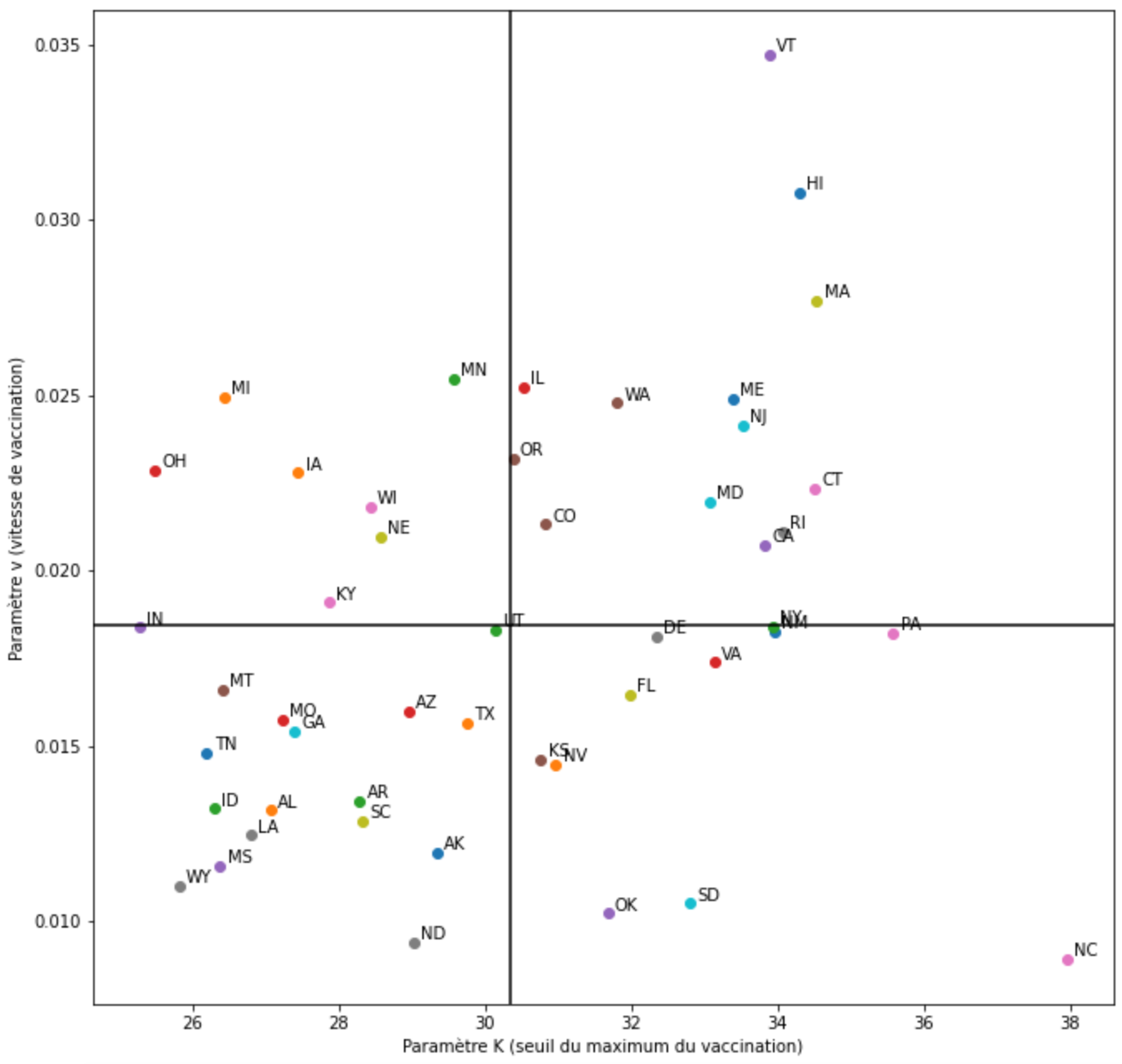}
\end{center}
\begin{center}
    \textit{Figure 3 : Distribution of K and $\nu$ according to the states, with the horizontal and vertical separations corresponding to the means}
\end{center}
\begin{center}
    \includegraphics[width=14cm]{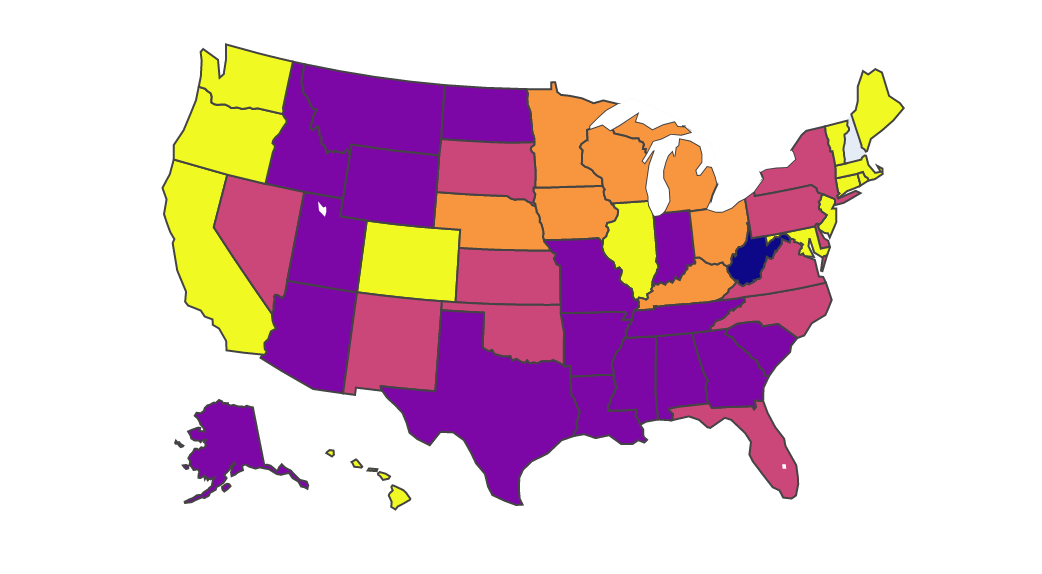}
\end{center}
\begin{center}
    \textit{Figure 4 : Graph of the distribution of K and $\nu$ in the United States}\\
    \textit{Legend : Yellow = Big K and Big $\nu$, Pink = Big K and Small $\nu$, Orange = Small K and Big $\nu$, Purple = Small K and Small $\nu$, Blue = No information }
\end{center}
\indent Very simply, if we look at the disposition of the parameters across the different states, we can see that it is the most developed states on the coast (California or Massachusetts) that have a population that has been vaccinated rapidly and in large proportions, and in contrast, it is the states in the center, such as Missouri or Arkansas, that have the lowest parameters in terms of vaccination rates. However, it is important to understand the correlation with the vulnerability index presented above. 

\subsection{Correlation between CCVI and vaccination parameters}

\indent At first glance, one would expect that, in general, the correlation between the vulnerability index and the ability to be fully and rapidly vaccinated for a county is negative. The larger the index, the more inequalities of any kind are present in the populations. However, in our framework for studying incentives, the goal is to optimize incentives, and thus to further study which of the six vulnerability characteristics are most likely to increase a county's overall vaccination rate.

\begin{center}
    \begin{tabular}{ |p{6.5cm}||p{3cm}|p{3cm}|  }
    \hline
    \hline
    \multicolumn{3}{c}{\textbf{Results table}}
    \\
    \hline
    \hline
    \textbf{CCVI} & K & $\nu$ \\
    \hline
    \textbf{Theme 1: Socio-economic status} & Slope = -4,61 \newline Corr = -0,084 &  Slope = -0,02  \newline Corr = -0,378\\
    \hline
    \textbf{Theme 2: Household composition and disability} & Slope = -12,42 \newline Corr = -0,221 &  Slope = -0,013  \newline Corr = -0,238\\
    \hline
    \textbf{Theme 3: Status and language of minorities} & Slope = 17,38 \newline Corr = 0,292 &  Slope = -0,002  \newline Corr = -0,027\\
    \hline
    \textbf{Theme 4: Type of accommodation and transportation} & Slope = -4,67 \newline Corr = -0,082 &  Slope = -0,005  \newline Corr = -0,087\\
    \hline
    \textbf{Theme 5: Epidemiological factors} & Slope = -1,50 \newline Corr = -0,029 &  Slope = -0,009  \newline Corr = -0,179\\
    \hline
    \textbf{Theme 6: Health System Factors} & Slope = -1,201 \newline Corr = -0,022 &  Slope = -0,012  \newline Corr = -0,236\\
    \hline
    \textbf{Global CCVI} & Slope = -2,75 \newline Corr = -0,049 &  Slope = -0,018  \newline Corr = -0,348\\
    \hline
    
    \end{tabular}    
\end{center}

In the same way as with a table, we can also draw a moustache diagram to get a better idea of the evolution of the K and $\nu$ parameters on average and in terms of standard deviation by dividing the counties into 10 compartments according to their distribution in each index.

\begin{center}
    \includegraphics[width=15cm]{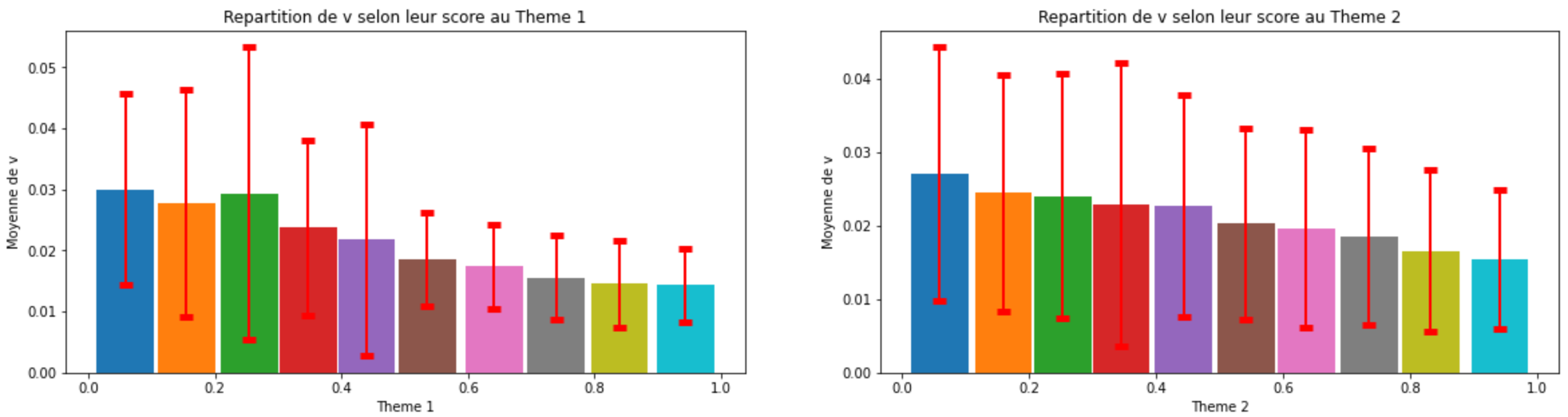}
\end{center}

\begin{center}
    \includegraphics[width=15cm]{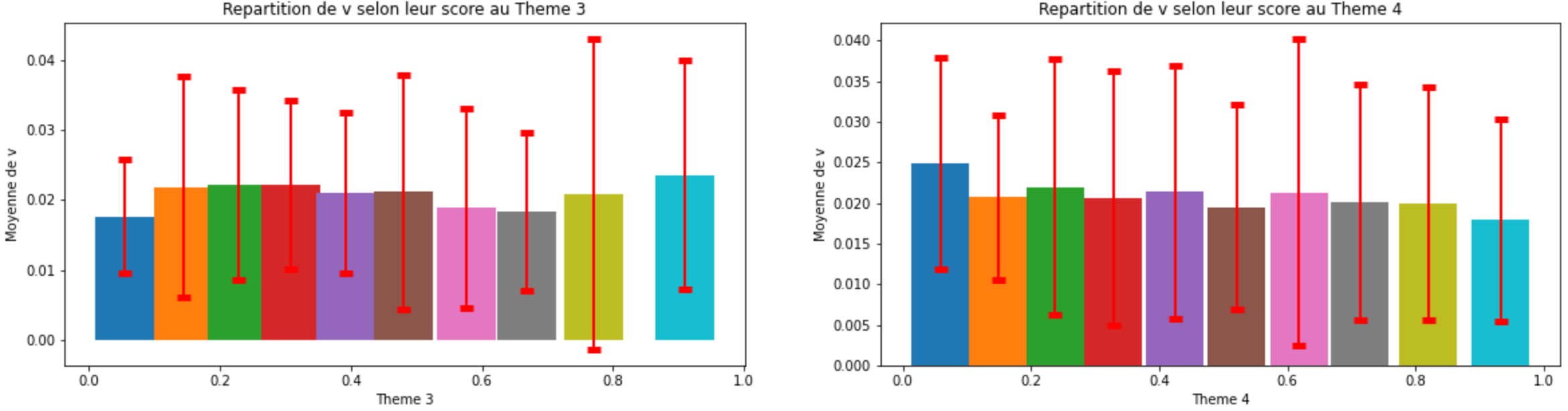}
\end{center}

\begin{center}
    \includegraphics[width=15cm]{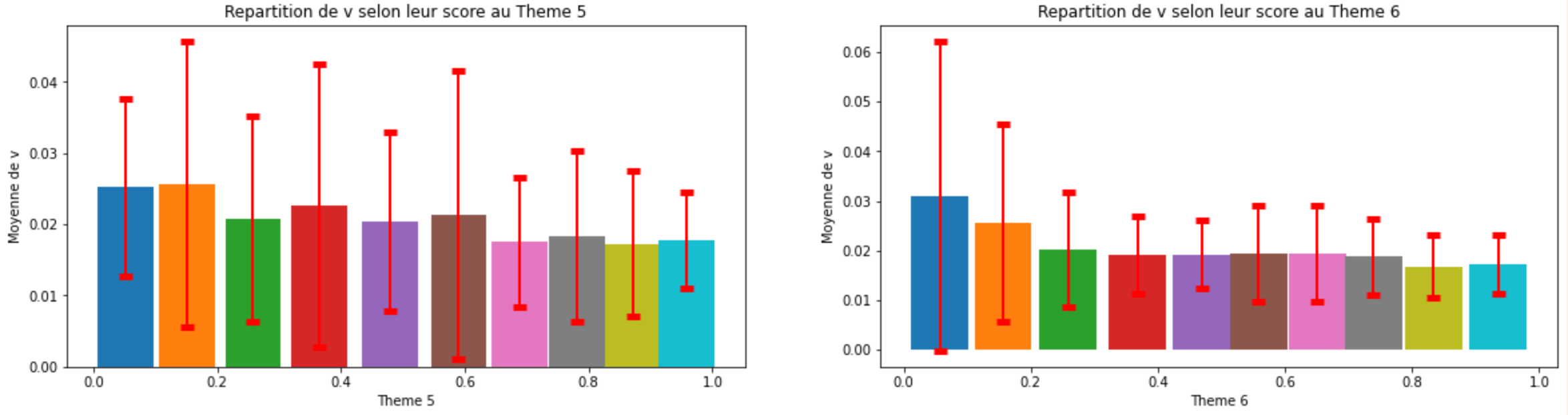}
\end{center}
\begin{center}
    \textit{Figure 5 : Mean and Standard Deviation of $\nu$ by the 6 CCVI themes and ranked by quartile}
\end{center}

\begin{center}
    \includegraphics[width=15cm]{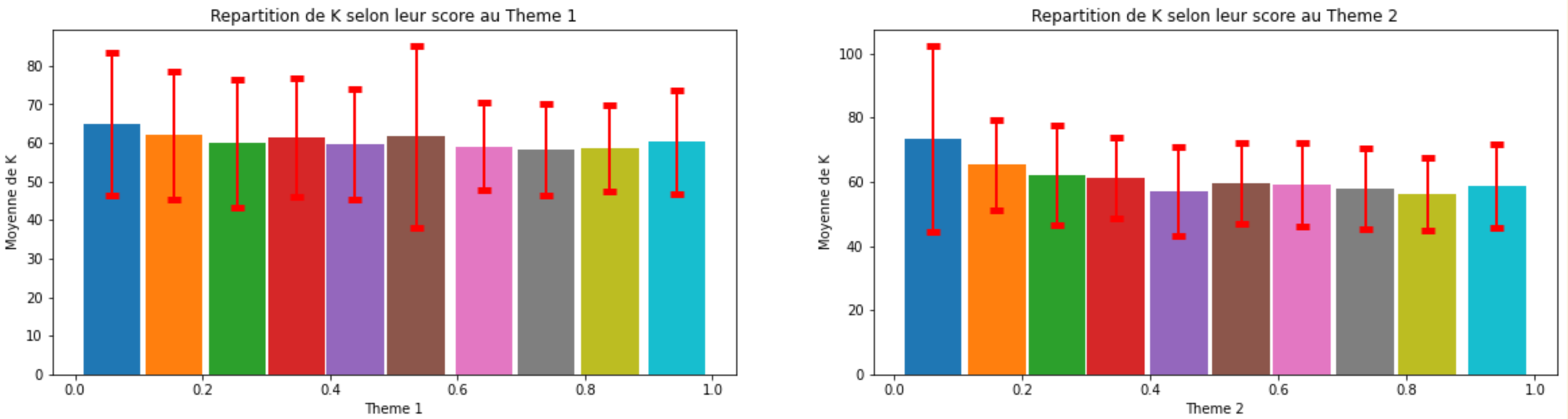}
\end{center}

\begin{center}
    \includegraphics[width=15cm]{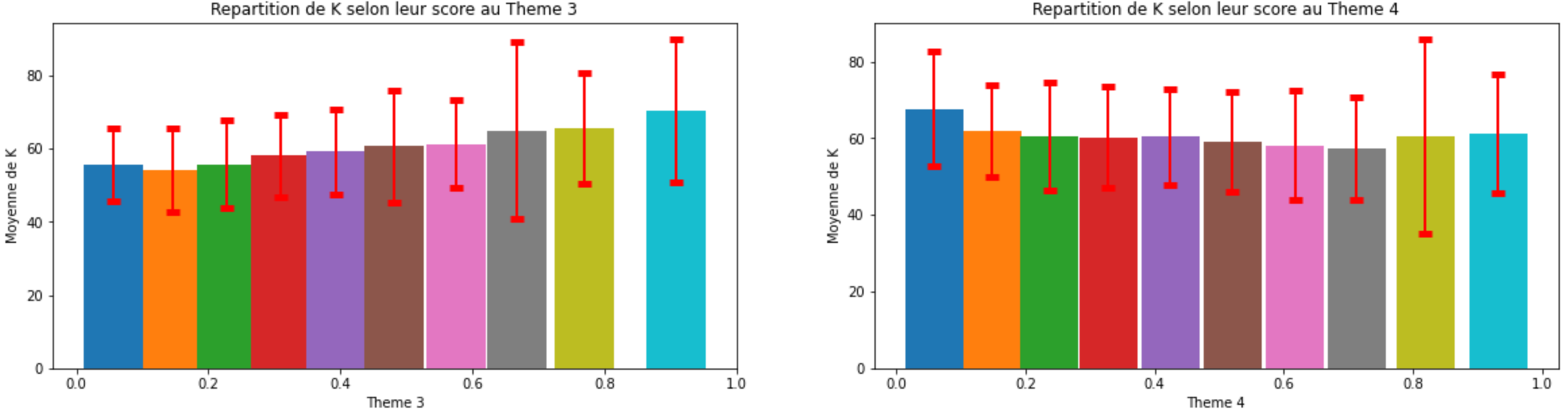}
\end{center}

\begin{center}
    \includegraphics[width=15cm]{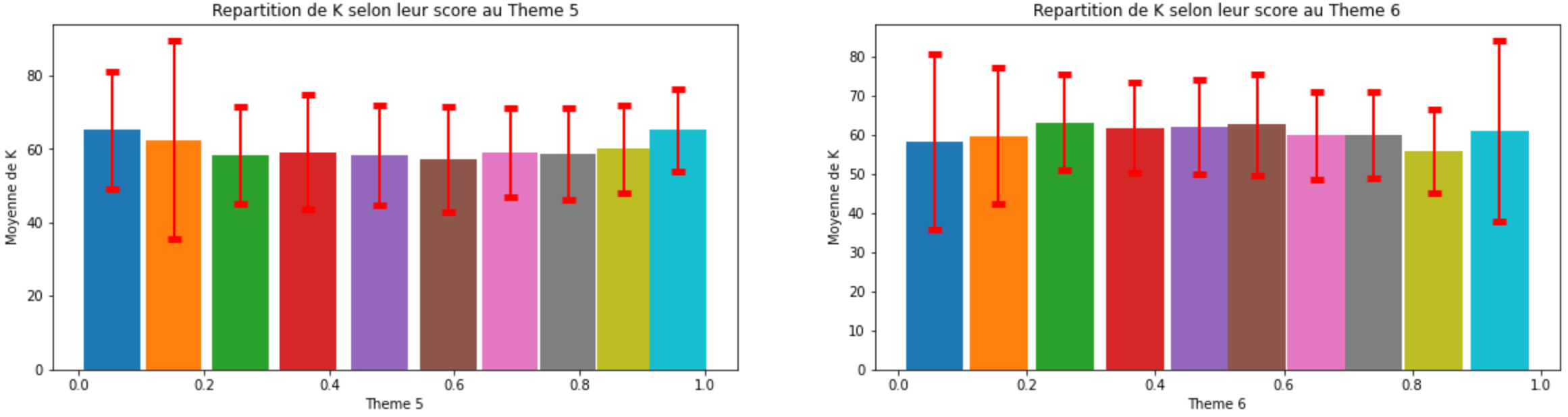}
\end{center}
\begin{center}
    \textit{Figure 6 : Mean and Standard Deviation of K by the 6 CCVI themes and ranked by quartile}
\end{center}

\begin{center}
    \includegraphics[width=15cm]{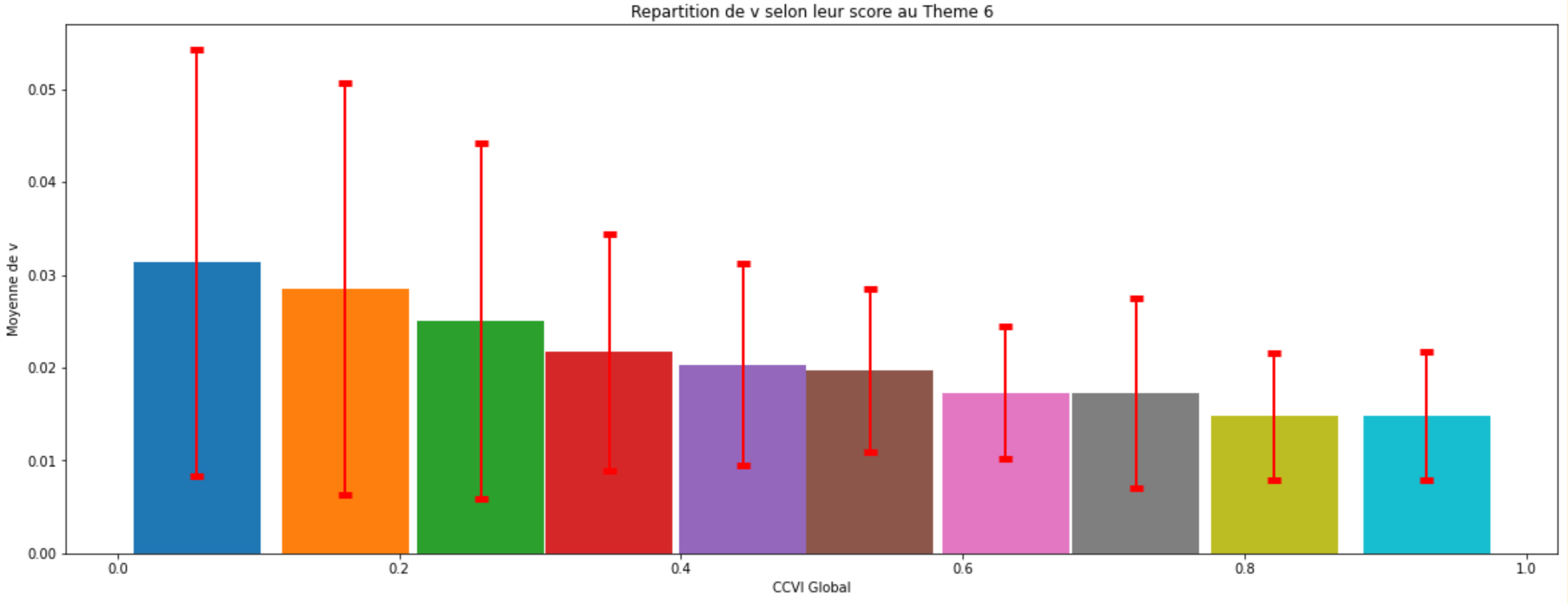}
\end{center}

\begin{center}
    \includegraphics[width=15cm]{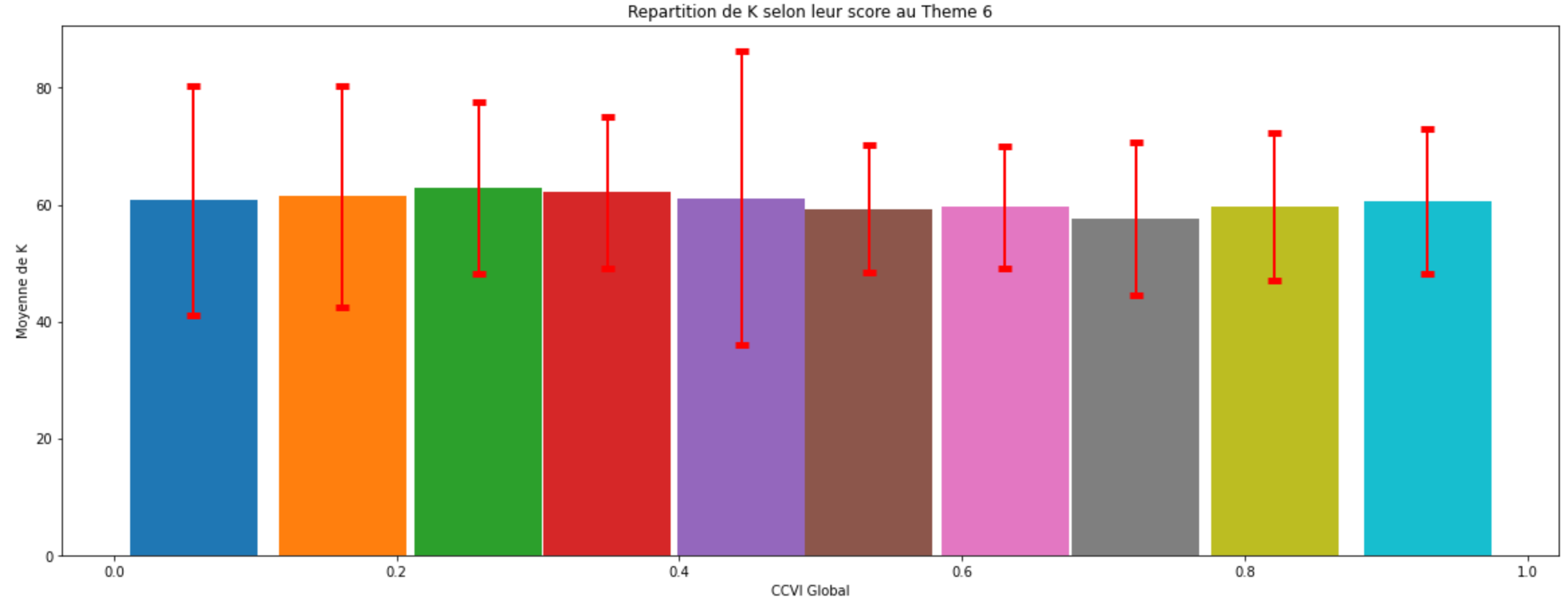}
\end{center}
\begin{center}
    \textit{Figure 7 : Mean and Standard Deviation of $\nu$ and K for the overall CCVI and ranked by quartiles}
\end{center}

\subsection{Results Analysis}

\subsubsection{General Conclusion}

Thus, through most of the visualizations presented above, we conclude as expected that, for both the maximum vaccination threshold and the speed, the correlation with the overall vulnerability index is on average negative for all US counties. It is also worth noting that, in terms of magnitude, it is mainly the speed of vaccination that is affected by populations with the greatest socio-demographic disparities and inequalities. This may be interpreted as not so much a lack of willingness of the more disadvantaged populations to be vaccinated but more a matter of accessibility and opportunity. 

\subsubsection{Targeted study by themes}

As far as themes are concerned, we can thus distinguish more specific populations if we choose to target rather to increase our K parameter or our $\nu$.\newline
\indent For the maximum threshold, we see that it is mainly the composition of households and situations of disability that can play a role. Thus, it seems interesting to set up measures that would seek to limit their field of action to large families, or to people with disabilities. In contrast, there is also a strong and positive correlation with regard to theme 3, the status and language of minorities. This suggests that it is not the reduction of this factor that makes the increase of the maximum threshold decisive. \newline
\indent For velocity, we first notice that the same themes are not at play. First, with a higher overall correlation than for K, we see that $\nu$ is negatively correlated mainly by socio-economic status, but also by household composition and health system factors in a second stage. It is thus economic inequality that accentuates inequalities in the speed of immunization of populations. 

\subsubsection{Incentive measures}

Finally, it should be noted that it is mainly the $\nu$ parameter that can be increased by reducing socio-economic inequalities. Thus, providing incentives to improve the standard of living of the poorest populations in certain counties, for example by taking up the measures proposing scholarships or even financial aid, seems to be a good solution to increase the vaccination rate in the United States. One can also imagine that each state has an interest in targeting counties where economic inequalities are high, in order to improve the efficiency of their measures and to avoid unnecessary costs in counties where the vaccination rate is already quite high.

\newpage

\section{Generalization of the model to counties and categorization by age group and immunization schedule}

\subsection{Motivation}

We have previously studied the impact of incentives at the state level. However, just as there are socio-economic disparities between states, there are also large socio-economic disparities between counties within a state. Since the incentive is the same for all counties in the same state, it is interesting to apply the synthetic control method to all counties to see if the vaccination patterns are different for the same incentive. 
\\

Similarly, stratification by age category is relevant. For example, in Delaware, the state offered a college scholarship to a randomly selected 12- to 17-year-old if he or she received a vaccine. For example, in Delaware, the state offered a college scholarship to a randomly selected 12- to 17-year-old if he or she was vaccinated. It is easy to imagine that this measure will not encourage adults to be vaccinated, since they are not affected. Such stratification also allows us to study the differential impact of incentives according to age category. 

\subsection{State of the art}

The use of county-stratified synthetic control has been done before to evaluate the effectiveness of vaccination incentives. In particular, the researchers studied the case of Ohio, which was the first state to implement such a measure through a lottery. E. Brehm et al. used the method developed by Abadie to evaluate the "cost" of a vaccination. They applied the synthetic control method to each county in Ohio, using counties in neighboring states (Indiana, Michigan and Pennsylvania) as control counties. Their study shows a positive effect on vaccination rates. Dividing the cost of the lottery (\$1 million) by the estimated number of additional vaccinations obtained through the lottery, they arrive at a cost of 75\$ per vaccination. Their stratification by county also allows them to study the disparity in behavior by county. Indeed, they observe a heterogeneity of behavior between the most populated counties, which have more vaccinations, and the least populated counties. Several explanations are put forward, such as the number of vaccination centers or a better access to information about the lottery. The robustness of their results is highlighted by the use of \textit{placebo tests} that confirm this trend. The researchers also sought to estimate the different responses by age category, but the lack of data on vaccination by age category was limiting.

\subsection{Data}

The data used in the remainder of this study come from several different sources, and have been the subject of research papers on which we have relied heavily. In order to combine these different datasets, the main resource used is the Federal Information Processing Standard (FIPS), a code that uniquely identifies a given county in the United States of America. A county is defined as a form of local government, a territorial division smaller than a state but larger than a city or town, within a state or territory. The total number of U.S. counties (or equivalents) in 2006 was 3,141, but our study will include fewer for reasons that will be discussed below.

\subsubsection{Immunization data by county}

The county-specific vaccination data used are available at \href{http://www.overleaf.com}{this address}, and are from the CDC (Center for Disease Control and Prevention), as of 5/19/2022. This dataset includes all vaccination data for U.S. counties from 12/13/2020 to 5/18/2022. It includes vaccination rates and raw data for different vaccine regimens, from 1st dose to full regimen, as well as stratification by age groups (12+, 18+, 65+) and census data for each county. 

To get an idea of the vaccine curves, some can be displayed (in the rest of the section, curves of vaccination rates for the 1st dose and the overall population will be displayed):

\begin{figure}[h]
    \centering
    \includegraphics[scale = 0.6]{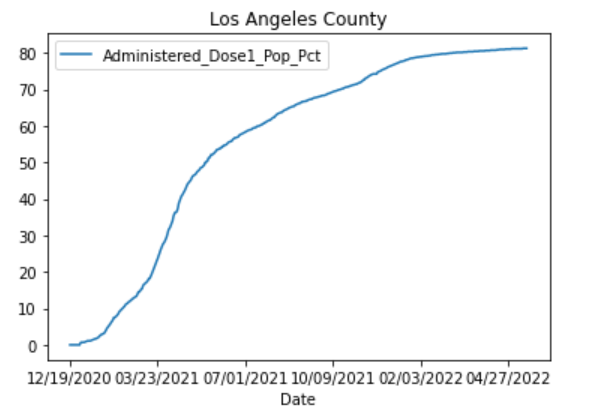}
    \captionsetup{labelformat=empty}
    \caption{Los Angeles County Vaccine Curve}
\end{figure}

We note that the shape of the vaccination curve is that of a logistic model. Nevertheless, some counties suffer from errors in the reporting of vaccination data, and there are two main types of errors:

\begin{itemize}
    \item[$\bullet$] Errors that contradict the cumulative nature of the immunization curve, with immunization rates decreasing over certain very short periods of time or being absent (NaN in the dataset)
    \item[$\bullet$] Carry-over errors that affect the immunization curve in an ad hoc manner
\end{itemize}

Two counties affected by errors of the first type can be represented: 

\begin{figure}[h]
    \begin{minipage}[c]{.46\linewidth}
        \centering
        \includegraphics[scale = 0.6]{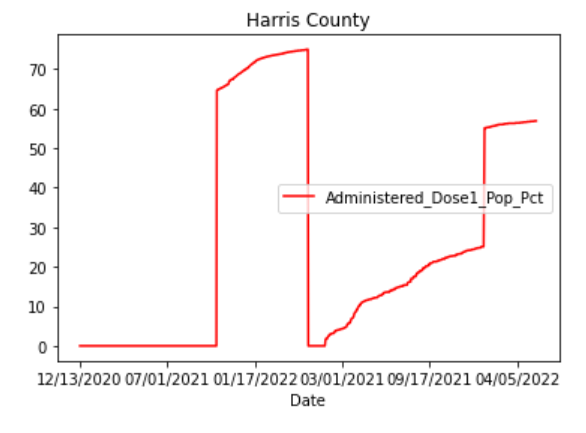}
        \captionsetup{labelformat=empty}
        \caption{Harris County Vaccine Curve}
    \end{minipage}
    \hfill%
    \begin{minipage}[c]{.46\linewidth}
        \centering
        \includegraphics[scale = 0.6]{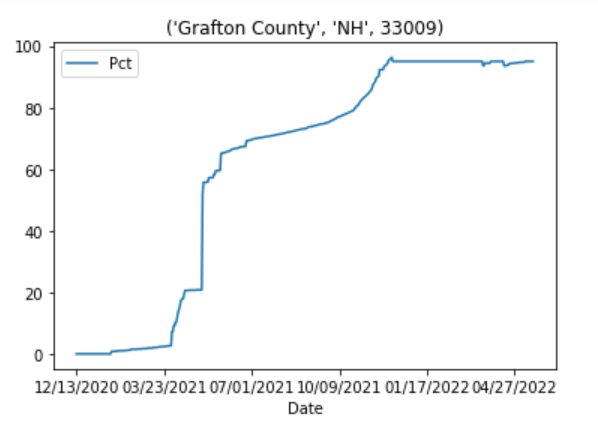}
        \captionsetup{labelformat=empty}
        \caption{Grafton County Vaccine Curve}
    \end{minipage}
\end{figure}

For the first county, we see that applying the cumulative maximum to the data would be sufficient to make them consistent. Nevertheless, we intuitively see that it is not normal to have this kind of data, with a zero vaccination rate for such a long period of time, and data that appear temporally inverted. There are several counties with such errors, and we made the decision for the most problematic ones to remove them from the dataset, when too many zero or "NaN" values appeared. For the second, the vaccine curve decreases and then exceeds the previous values. For this type of condition, the periods over which carryover errors are found being very small in comparison with the total duration of the study (of the order of 1\%), we decided to linearly interpolate the missing values. Here are the curves after data processing: 

\begin{figure}[h]
    \begin{minipage}[c]{.46\linewidth}
        \centering
        \includegraphics[scale = 0.6]{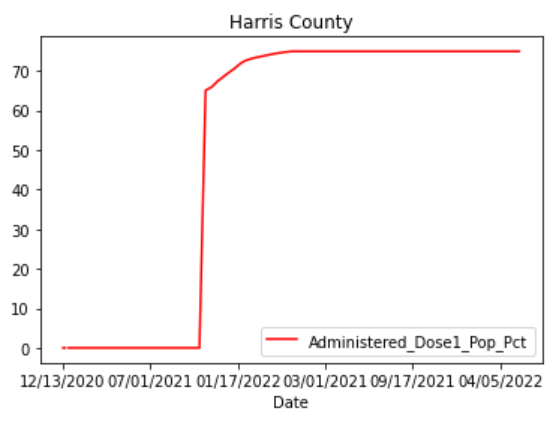}
        \captionsetup{labelformat=empty}
        \caption{Harris County Vaccine Curve}
    \end{minipage}
    \hfill%
    \begin{minipage}[c]{.46\linewidth}
        \centering
        \includegraphics[scale = 0.6]{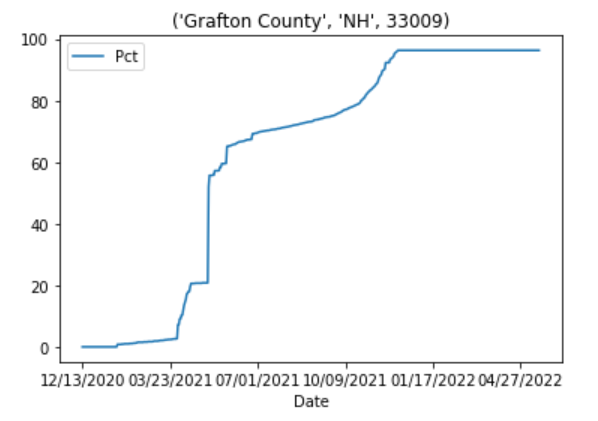}
        \captionsetup{labelformat=empty}
        \caption{Grafton County Vaccine Curve}
    \end{minipage}
\end{figure}

The second type of error is easily understood from a visualization:

\begin{figure}[h]
    \centering
    \includegraphics[scale = 0.7]{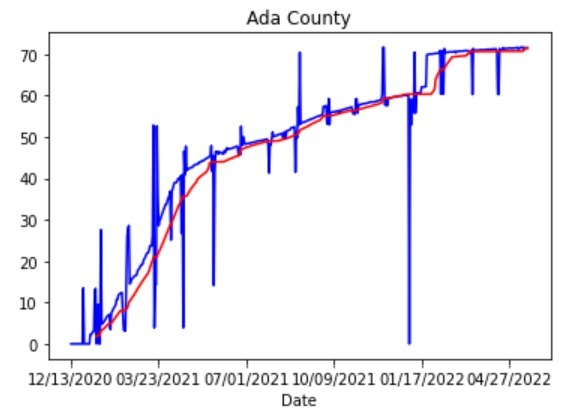}
    \captionsetup{labelformat=empty}
    \caption{Vaccine curve (smoothed in red) for Ada County}
\end{figure}

The data are highly noisy, and we decided to apply a 7-day rolling average, in order to remove the fluctuations related to data carryover errors so as to highlight the real vaccination trend, which allows us to obtain the red curve.\newline

Finally, in order to obtain rates for a new age category (e.g., 18-65 from 18+ and 65+), we were able to directly exploit the raw immunization data contained in the dataset, ensuring that when carryover errors were present for one age category and not another, the errors did not propagate along the new age category created.

\subsubsection{Study parameters by county}

The data used for county parameterization were finely selected based on two studies from which we drew. The first is from the Journal of Immigrant and Minority Health [6], and the dataset was described in 3.2 and presents the different themes of the Community Vulnerability Index (CVI). These different themes were defined by the CDC and the Surgo Foundation and these indices have been used in many studies investigating the impacts of COVID-19 in American communities. The second is from an article called \textit{Risk factors for increased COVID-19 case-fatality in the United States} [7], which seeks to identify county-level variables associated with COVID-19 case-fatality. Their goal is to pool factors responsible for case-fatality rates and identify the most relevant ones in order to help scientific studies related to this topic such as ours. The different risk factors are grouped under the following categories: demographic, socioeconomic, access to care, prevalence of comorbidities and non-pharmaceutical interventions.\newline
A correlation and significance analysis was performed in this paper, which allowed us to discard the most correlated parameters or those with a p-value too low for the purpose of this paper, and to take into account only those that could be significant for our model. We therefore selected, in addition to the CCVI parameters, the following 6 parameters for their significance, their high variance, or their lack of correlation with the previous parameters, as well as for qualitative and computational criteria:

\begin{itemize}
    \item Stroke Hospitalization Rates for Medicare Beneficiaries
    \item Hospitalization rates for hypertension among Medicare beneficiaries    
    \item Hospitalization rates for all cardiac diseases among Medicare beneficiaries
    \item Prevalence of heart disease among Medicare beneficiaries
    \item Mortality rate of hypertension in the 35+ age group
    \item Mortality rate for cardiovascular diseases in the 35+ age group
\end{itemize}

The correlation matrix of the parameters can be displayed:

\begin{figure}[h!]
    \includegraphics[scale = 0.7]{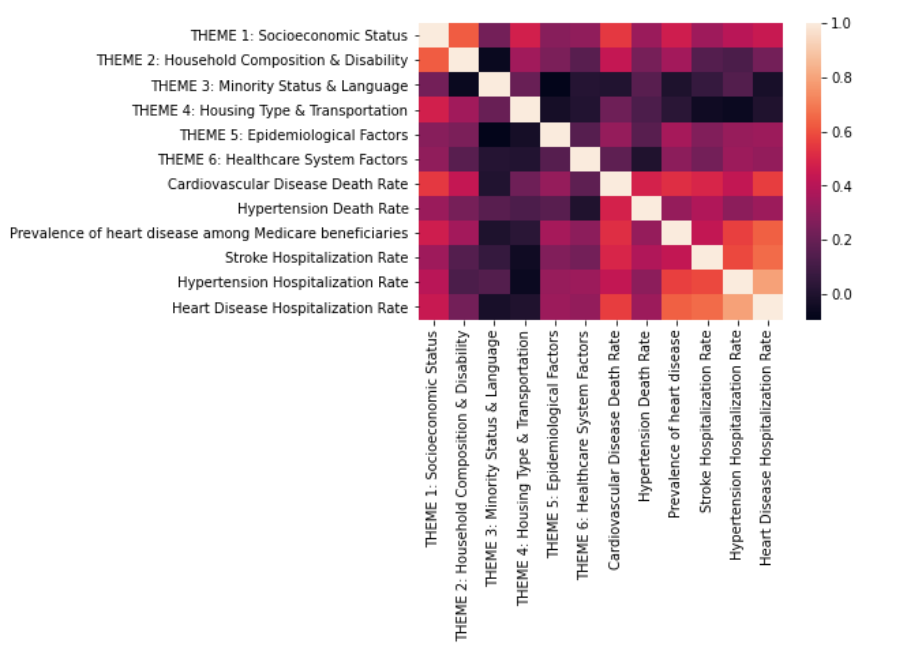}
    \captionsetup{labelformat=empty}
    \caption{Correlation matrix of the parameters}
\end{figure}

\subsubsection{County Clustering by the American Communities Project}

The \textit{American Communities Project} \textbf{[8]} is an initiative of the State Michigan University School of Journalism to group American counties into 15 clusters according to a panel of 36 socio-demographic factors. The method used is a classic clustering method. The goal of this initiative is to better understand "\textit{the different types of communities that make up America so that new and better ways of recognizing and evaluating what works and what doesn't can be developed}". The table below lists the 15 clusters and their corresponding number of counties (we leave the English names here because some names are not translatable into French): 

\begin{table}[h]
\centering
\begin{tabular}{ | c || c | c | }
\hline
    Category & Number of counties & Proportion \\
    \hline \hline
   Exurbs & 222 & 7,1 \% \\
    Graying America & 364 & 11,2 \% \\
    African American South & 370 & 11,8 \% \\
    Evangelical Hubs & 372 & 11,8 \% \\
    Working Class Country & 337 & 10,7 \% \\
    Military Posts & 89 & 2,8 \% \\
    Urban Suburbs & 106 & 3,3 \% \\
    Hispanic Centers & 161 & 5,12 \% \\
    Native American Lands & 43 & 1,37\\ 
    Rural American Lands & 599 & 19,06 \% \\
    College Towns & 154 & 4,9 \% \\
    LDS Enclaves & 41 & 1,3 \% \\
    Aging Farmlands & 161 & 5,1 \% \\
    Big Cities & 47 & 1,5 \% \\
    Middle Suburbs & 77 & 2,4 \% \\
   \hline

 \end{tabular}
    \caption{List of the 15 clusters of the \textit{American Communities Project}}

 \end{table}

This clustering is similar to the one performed during the first semester on our states, but it is more refined. It will especially allow us to reduce the number of control counties as we will see in the following. 

\begin{figure}[h!]
    \centering
    \includegraphics[scale = 0.2]{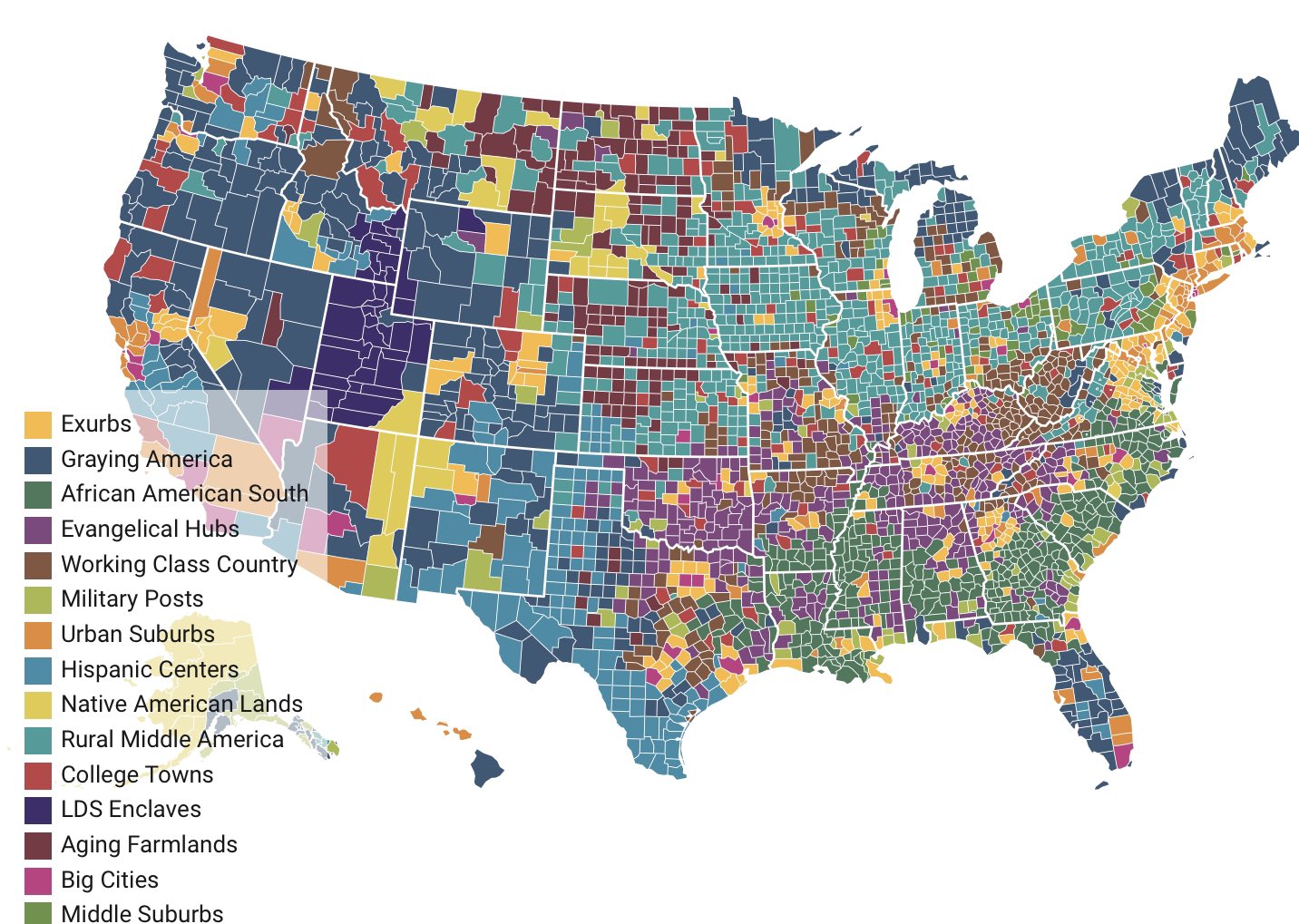}
    \captionsetup{labelformat=empty}
    \caption{Map showing the clusters of the\textit{American Communities Project} }
\end{figure}

\subsection{Implementation}

In order to perform the  \textit{synthetic control}on counties and by age category, we have used the same architecture as the code used for the \textit{synthetic control} by states. Nevertheless, the handling of several complex databases required many modifications detailed here. 

\subsubsection{Problems encountered}

The previous sub-section detailed the main problem of databases concerning the lack of data or the presence of outliers. We thus had to use a smoothing function. 

Another problem encountered is that of correspondence between the different databases. Several problems were encountered:

\begin{itemize}
    \item[$\bullet$] Some databases did not include all counties. Since we needed to have every county represented in each of our databases, if a county was not present in one database, we had to remove it from all the others. 
    \item[$\bullet$] The names of the counties differ between the different databases. Moreover, two different states can have a county with the same name. So we made the joins not on the county names, but on the FIPS code.

\end{itemize}

These manipulations slightly reduced the number of counties processed. From 3141 counties processed, we have gone to 3131.

\subsubsection{Restriction of the database by age category and immunization schedule}

The CDC vaccination database is a very rich database stratified by age category. We have chosen to code a function \textit{cat age vax(lb,ub,vax scheme)} which lets the user choose the desired stratification. The arguments of this function are the limits of the desired age category as well as the desired vaccination scheme (1st dose, complete vaccination). According to the chosen stratification, this function cuts the dataframe to keep only the data used by the \textit{synthetic control}. \newline
\noindent
For example, the following command : 

\begin{center}
    \includegraphics[scale = 1]{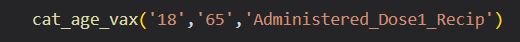}
\end{center}
\noindent
allows us to retrieve 1st dose data for 18-65 year old. 

\subsubsection{Data smoothing function}

As mentioned in the section on immunization data, immunization data may require temporal smoothing. We therefore first apply a function that removes the county if it contains too many anomalies (too many 0s or NaNs over the total duration), then an interpolation function when the time periods where data are missing are very small in relation to the total duration (of the order of a few days out of a total of 522 days) and a sliding window of one week in order to average the curve and eliminate fluctuations that could be due to a data carryover error, 

\subsubsection{Synthetic control}

Let us now turn our attention to the implementation of the \textit{synthetic control}. As with the states, we had to separate the so-called control counties that did not experience incentives from the target counties. To do this, we simply separated the counties in all control states from the counties in the target states. This represents 2120 control counties and 1013 target counties. Also, as explained earlier, we focused only on the target counties in Ohio. Since the computation time was very large for one county, we could not run the \textit{synthetic control for all counties}. Therefore, we focused only on the Ohio counties. This choice is motivated by the relatively good data we have for the counties of this state but also because many research papers deal with this state. This will allow us to compare our approaches. The high computational time is also due to the large number of control counties. We have therefore established different strategies to reduce the number of control counties while trying to minimize the selection bias. These strategies are detailed in the next paragraph.

\subsubsection{Strategies for reducing the number of control counties}

\paragraph{Regularization}

The first mathematical strategy implemented to reduce the number of control counties during the optimization phase and to make the weight vectors more \textit{sparse} was regularization, as presented in the corresponding section of the study for states. We therefore fixed the coefficients as in that section, but when the number of control counties was too high, the weights would get squashed and standardized in inverse to the number of control counties. We then opted for a different approach, which is detailed in the next paragraph.

\paragraph{Initial weights of control counties and parameters in the synthetic control}

The optimization procedure within synthetic control requires starting from an initial point for the parameter and control county weights. One can set, as in some studies on synthetic control, the initial weights of the parameters to the inverse of the variance of the corresponding parameter, and launch the optimization from this initial point. Nevertheless, for the parameters considered in our study, the variances being very disparate as can be seen in the figure below, the weights of the CCVI parameters would have been strongly advantaged compared to the others. The corresponding fit results were found to be weaker than for a random initialization of the control county and parameter weights. \newline

\noindent
We also tried a uniform initialization of the weights, but this choice rendered in the vast majority of cases (on average 8 times out of 10 on 5 tests of samples of size 10) the optimization null and void, and the result returned was the initial value, resulting in a synthetic curve being simply the arithmetic mean of the control counties, and thus the same curve for all target states studied.\newline

\noindent
Once the initialization is done in a non-uniform random way, we launch the optimization procedure a first time, we set to 0 the weights that are too low (below half of the 5th most important state in the synthetic county, which leaves in general between 5 and 8 states), and we normalize before launching the optimization procedure again. This choice allows us to start with a vector of weights close to a local optimum and chosen in a non-biased way, because we base ourselves on the result of the optimization and the result returned is calculated and not fixed by us. We chose this approach to obtain sparse weight vectors, which we will present next.

\paragraph{Selection of counties by the American Communities Project}

Another tactic implemented to reduce the number of control counties is to rely on the county categories given by the \textit{American communities project} [8] [8]. Thus, for a given county, only counties in the same category are kept as control counties, which allows for socio-demographic stratification and greatly reduces computational time while providing weights for synthetic counties that are easily interpreted, given the similarity of the states.

\paragraph{Other possible tactics: geographical selection and from the synthetic control by state}

In the same way that one can select counties based on the stratification given by the American Communities project [8], one can rely on the control counties selection approach implemented by A. Abadie in [2], and preliminary select only the control counties located in the states neighboring the target counties considered. Nevertheless, this approach was invalidated because, although it could provide satisfactory results in terms of quality of \textit{fit}, the control counties considered in the linear combination of the target county do not necessarily share relevant socioeconomic characteristics or health criteria with the target county. \newline

For example, in the figure below, an attempt was made to model Hamilton County, Ohio, by the control counties of neighboring states (Pennsylvania and Indiana). This results in a combination of approximately 79\% by Dauphin County (PA) and 21\% by Pike County (IN). Nevertheless, based on quantitative criteria such as the stratification of [8], we see that Hamilton County is part of the \textit{Urban Suburbs}, with a dense and very diverse population, with a Democratic majority, while Dauphin County is categorized among the  \textit{College towns}, with a white student population (81\% on average for these counties) and Pike County is among the  \textit{Exurbs}, with a relatively well-to-do population, with a median household income of 77. 200\$ and little diversity, with a Republican majority. It therefore seems very inconsistent to model such a county from counties with socio-demographic characteristics that are not very close to this one, even if these counties are located in two neighboring states, given the vast size of the states and the diversity of the counties that make them up. This is what led us to prefer the approach of county selections by the method based on [8].\newline

Finally, another method of selecting control counties that we considered would have been to select only the counties in the most influential control states in the synthetic of the target state whose target counties we are trying to model. Nevertheless, this approach would introduce a selection bias that is difficult to quantify, which encouraged us to abandon it.

\begin{figure}[h!]
    \centering
    \includegraphics[scale = 0.5]{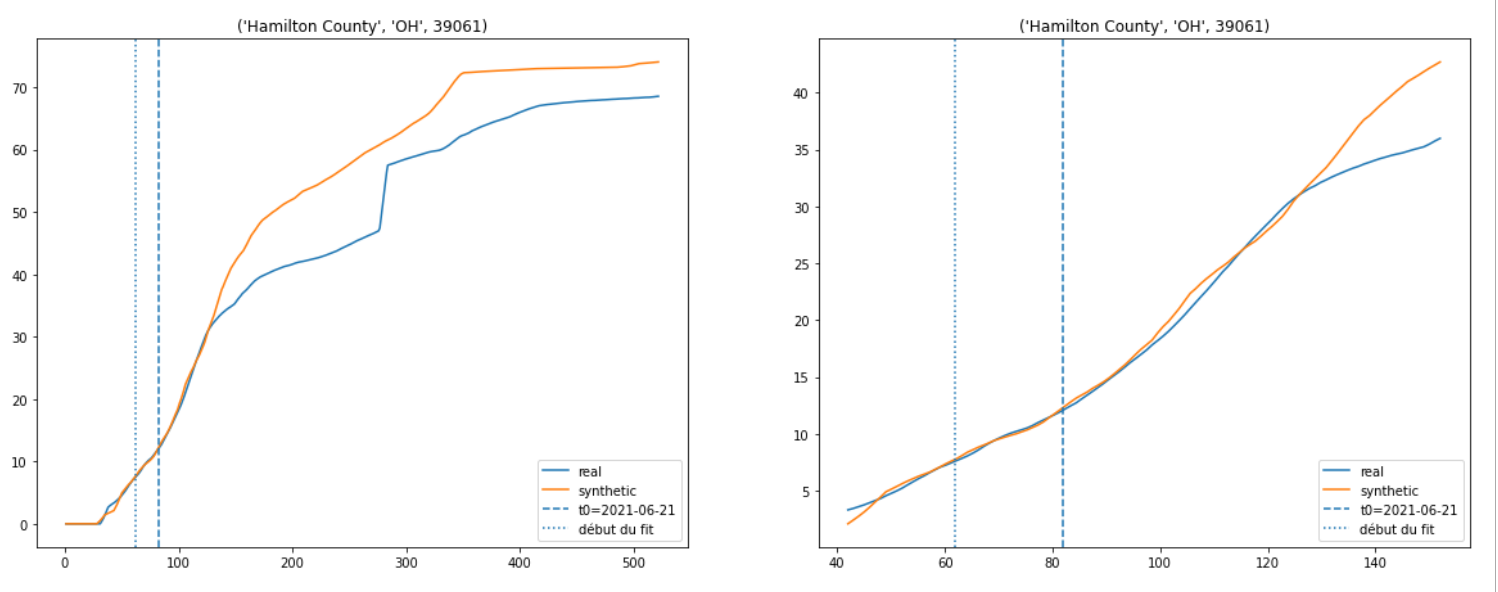}
    \captionsetup{labelformat=empty}
    \caption{Synthetic curve for Hamilton County from neighboring state control counties}
\end{figure}

\subsection{Analysis of the results}

This section analyzes the results of the synthetic immunization curves for counties. As specified above, we restricted ourselves to selected counties in Ohio. 

\subsubsection{Overall trend}

It is difficult to identify an overall trend since we did not conduct the synthetic control for all counties. Nevertheless, since we conducted the synthetic control for counties with different socio-demographic profiles, we can still draw some conclusions. For example, we found that the incentives were more effective overall in the most populous counties than in the moderately populous counties. In these counties, the incentives were only marginally effective overall. We detail in the next section curves of three counties whose fittings worked well. 

\subsubsection{Qualitative study of the behavior of selected counties}

To illustrate these overall observations, we will compare the results for three Ohio counties: Franklin County (the county that includes the metropolitan area of Columbus, the capital of Ohio), Hamilton County (the county also includes a large city, Cincinnati), and Montgomery County (the county with a smaller city, Dayton). 

\paragraph{Franklin County}
\begin{center}
\includegraphics[width=15cm]{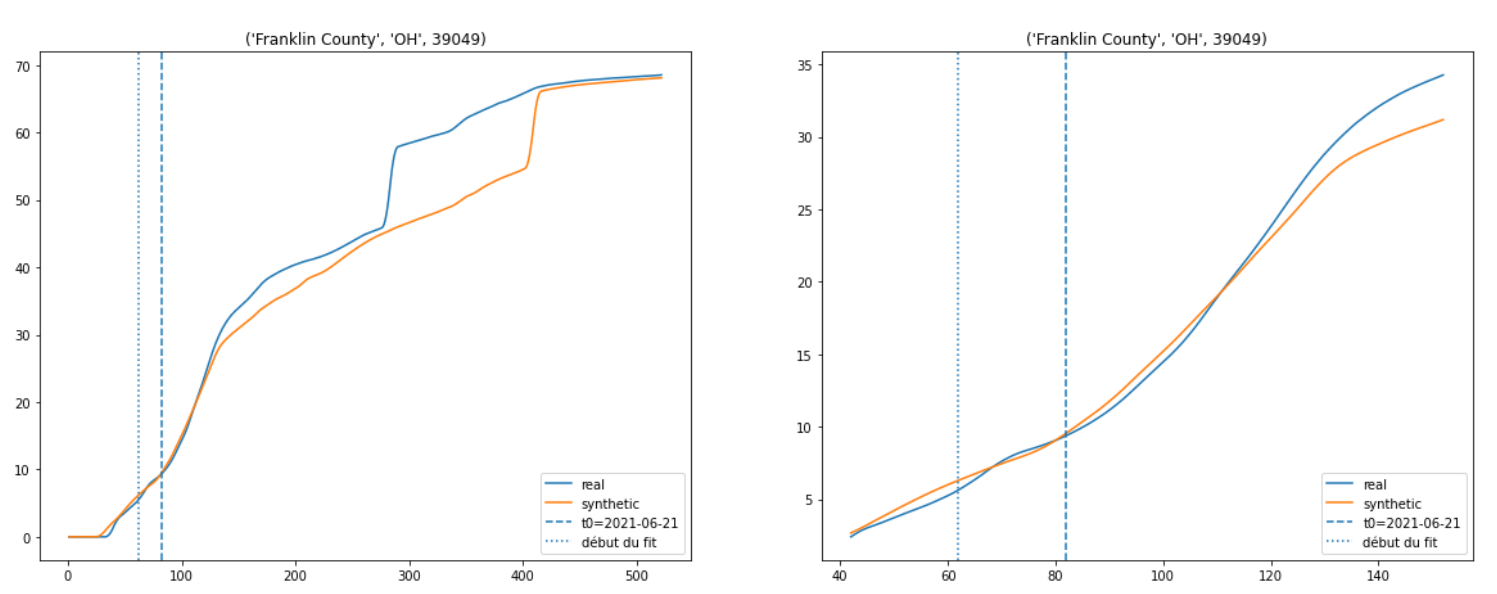}
\end{center}
\indent We observe a delayed effect of the incentive. Nevertheless, it seems to have a positive effect a few dozen days after its implementation.\\
However, it is difficult to ignore the big jump in the actual vaccination curve at about 280 days. This is a feature common to all Ohio counties and is related to the state's vaccine data collection..
\newpage
\paragraph{Hamilton County}
\begin{center}
\includegraphics[width=15cm]{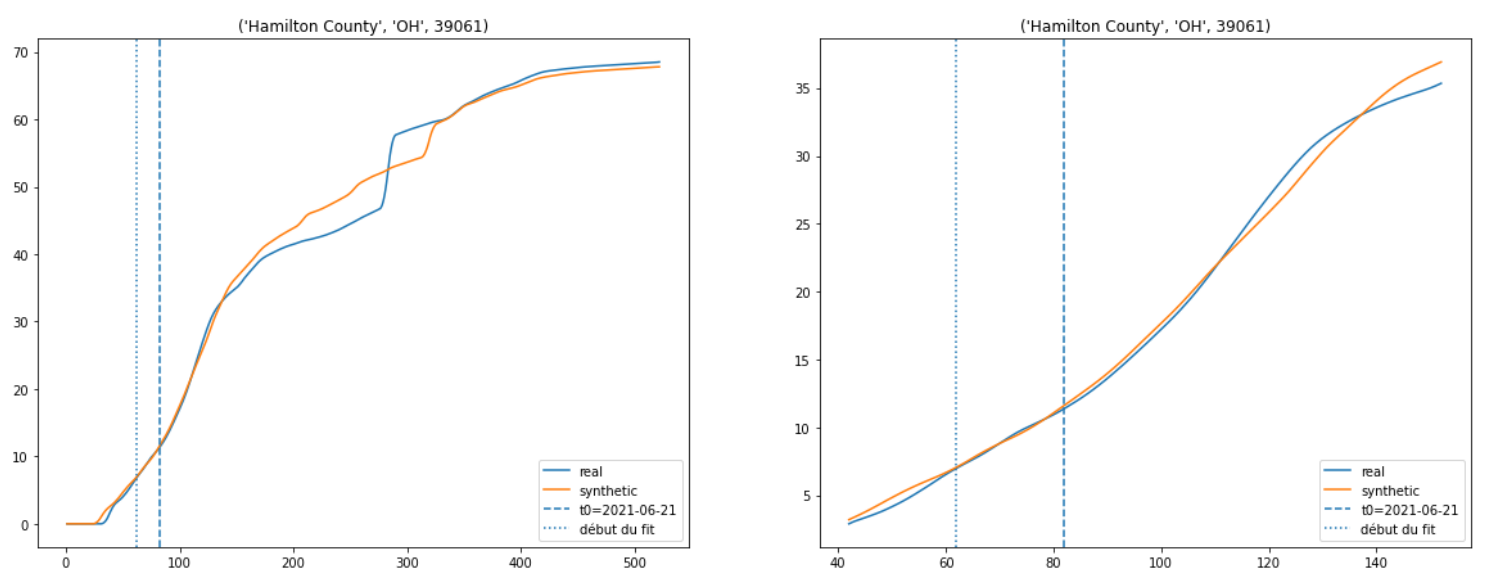}
\end{center}
\indent The incentive appears to have had a non-existent effect on Hamilton County. Indeed, we even observe the synthetic curve exceeding the actual curve after about 50 days.
\paragraph{Montgomery County}
\begin{center}
\includegraphics[width=15cm]{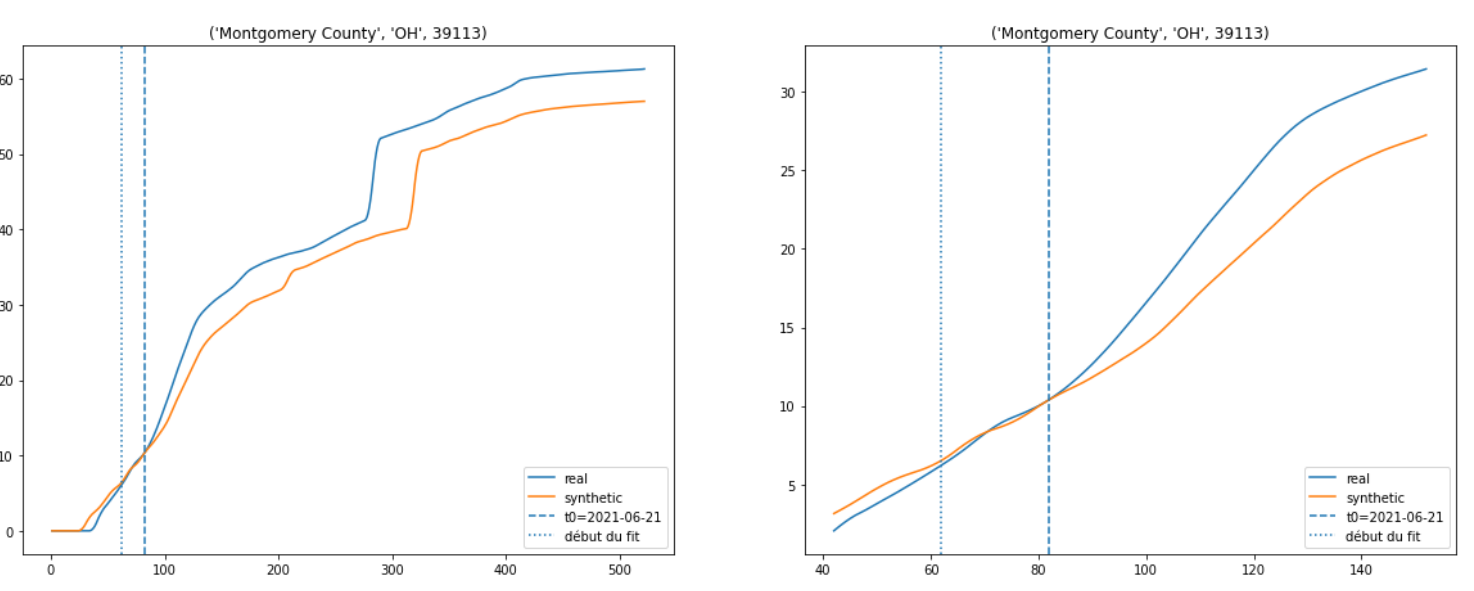}
\end{center}
\indent We see in Montgomery an immediate effect of the measure. It thus seems to have had a positive effect on vaccination. Nevertheless, one can question the reality of such a thing because it seems unlikely that the population would react with such a magnitude within a day to a government announcement. This type of information usually takes a few days for the population to act on it.
\\

\indent In conclusion, we observe that 2 out of 3 of these urban counties have seen positive effects from the incentives, which rather confirms the global conclusions preceding their analysis but encourages a more in-depth analysis on this point. It may also be added that the fit period is rather short. This does not seem to have affected the quality of the fit but makes these results somewhat less reliable.

\subsubsection{Robustness of results}

In order to assess the robustness of the results, \textit{placebo tests} would need to be implemented as for the states. The stratification of counties was also a motivation to improve the robustness of the results since the placebo tests require a significant amount of control entities to approximate a density, which we did not have per state given the small number of control states. With over 2000 control counties, the \textit{placebo tests} would have been of good quality. For example, the \textit{placebo tests} performed by E. Brehm et al. are shown below: 

\begin{figure}[h]
    \centering
    \includegraphics[scale = 1]{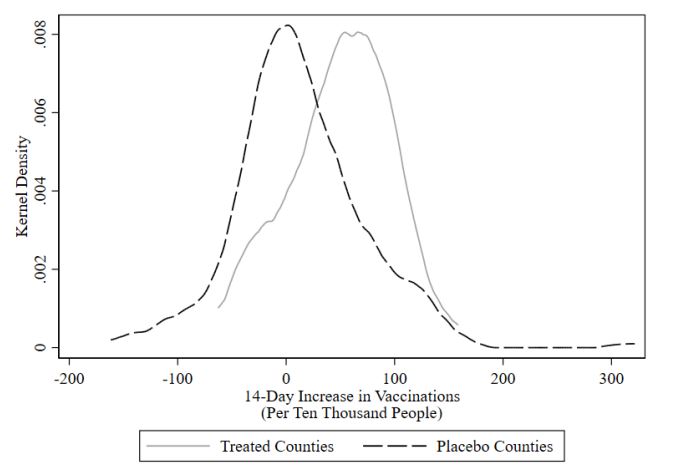}
    \captionsetup{labelformat=empty}
    \caption{Placebo Tests performed by E. Brehm et al.}
\end{figure}

Nevertheless, the \textit{placebo test} requires to realize the synthetic control for each control county, which would mean to run our code more than 2000 times. As the computation time is consequent, this part is in progress. 
\newpage 

\section{Graphic interface}
\subsection{Background and presentation of the dashboard}
The last part of the S8 project is the interactive dashboard. We had already built an interactive dashboard during the S7 project which was very efficient, allowing to visualize the results and then interpret them more easily. This interface was designed with Dashly, a Python package. \\ \\
\indent The motivation for building a new interface was twofold. First, the S7 version of the dashboard was designed primarily as a tool to accompany the report and to better visualize the results. The first motivation was to produce a dashboard that was not reserved only for those who had read and understood the report but rather could be used by "everyone", at least by public health decision-makers. The second goal was educational. We wanted to take advantage of the project to "get to grips" with data visualization tools and software that are widely used in the professional world. This second point provoked a total redesign of the interface to build it in the new environment, Tableau.

\begin{center}
    \includegraphics[width=15cm]{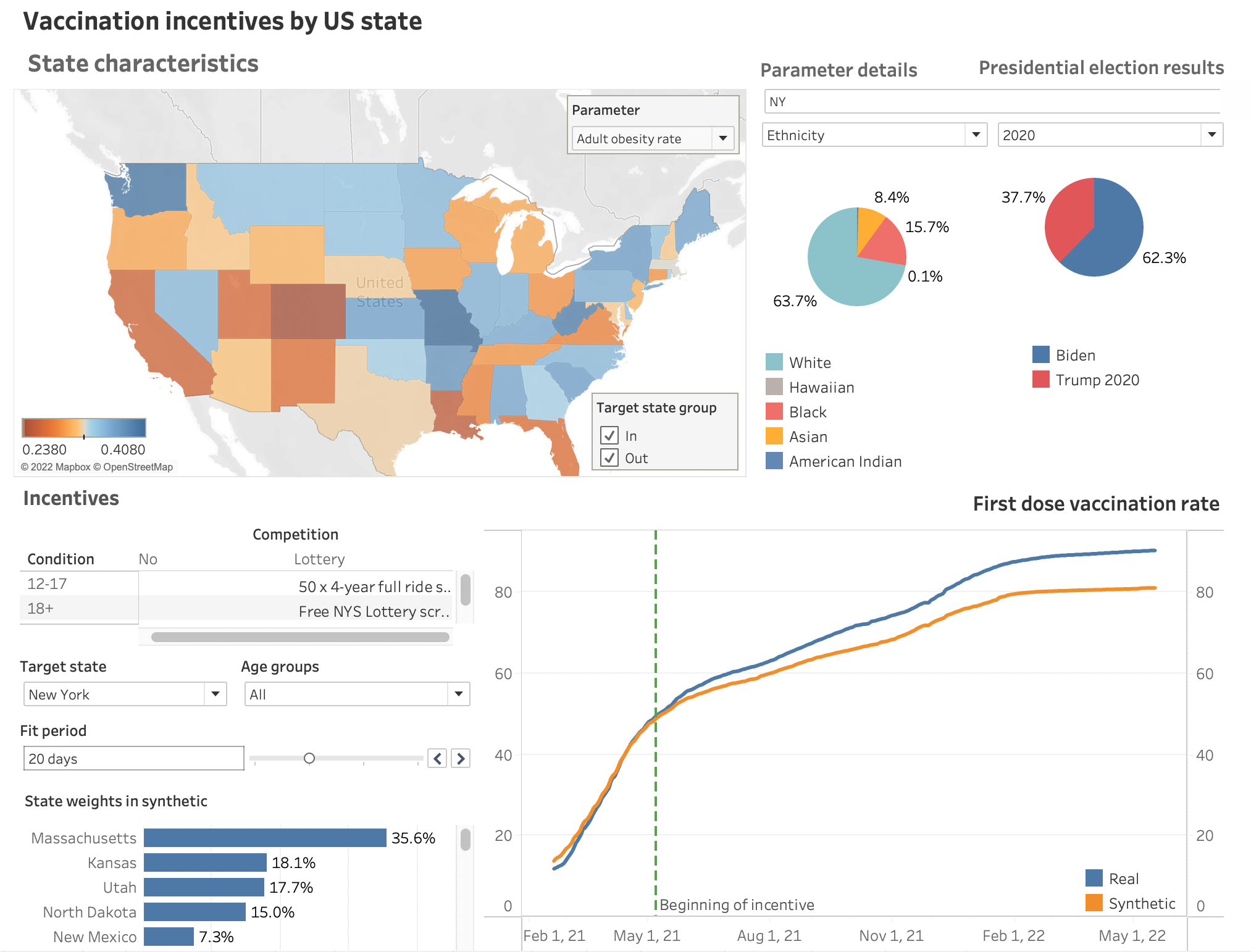}
\end{center}
\newpage
The interface can be deconstructed into 4 main areas:
\begin{itemize}
\item the card that prompts you to choose a state (top right)
\item the parameters of the selected state (top right) 
\item Vaccine curves for the selected state (bottom right)
\item the parameters of the synthetic control (bottom left) \\
\end{itemize}

\subsection{Use cases}
Here is an example of how the interface is used by a typical user from the public health sector:
\begin{enumerate}
\item The user wants to see in a particular state whether the incentives have resulted in an acceleration of immunization. They click on the state in question on the map.
\item At the bottom right, the user can observe the evolution of the synthetic curve and the real curve. He/she can also choose an age group to see the efficiency of an incentive specific to an age group.
\item He.She can observe the robustness of the synthetic model by looking at the stability of the curve for different duration of training period as well as the stability of the control states that model the state.
\item To put the nature of the incentive in context, the user can refer to the diagrams explaining the metrics grouped by theme at the top left of the screen.
\item Finally, if the user wants to compare the group of processed states with the group of control states, he/she can choose a parameter at the top right of the map and select and deselect in turn the processed states with the "In/Out" option at the bottom right of the map.
\item Or the user can select a new state to see the effect of the incentives and possibly compare it to the previous state.
\end{enumerate}

\newpage

\section{Opening}

\subsection{Continue the study by county}

Due to the considerable amount of computational time, we were not able to conduct our stratified county study in its entirety. We had to restrict ourselves to a small number of counties within Ohio. Eventually, we would like to perform the \textit{synthetic control} for all counties that have experienced an incentive. A study of the robustness of our results would also need to be conducted as we did for the states. Finally, since some vaccination data are of poor quality, we could further refine our smoothing function.

\subsection{Stratification by ethno-racial category}

As we have done with counties, and age categories, we would like to do a study stratified by ethno-racial category. It would be very interesting for governments to know how effective incentives have been by these categories in order to better target them in the future. In terms of data needs, this requires having all vaccination data by ethno-racial category by state (or even by county). However, these data are not centralized and most are not public. We had begun to collect all of these data, but it is necessary to approach each health center in each state individually in order to recover them when they are not public. So far, we have collected data for about ten states. Once this collection is completed, we will be able to compare our geographic, socio-demographic and ethno-racial stratifications. 

\subsection{Evolution of vaccination parameters (K and $\nu$)}

\indent As presented in Section 3, we were able to highlight the importance of certain inequality factors within populations that play a significant role in both the maximum vaccination rate threshold and the rate at which it is reached across US counties. However, in our analysis, the K and $\nu$ vaccination parameters are fixed in time. Thus, it would have been interesting to compare the evolution of these parameters before and after the implementation of the incentives, to see if their increase was significant, and especially to understand which vulnerability factors are the most elevated in the counties where K and $\nu$ have increased the most.

\subsection{Incentives in late 2021 and the 5-11 age group}

A second wave of incentives has been implemented by the states since late summer 2021 in states such as West Virginia with "Do it for Babydog Round 2" and "Do it for Babydog Round 3" or in California where the measures have been continued and improved until January 2022. \\ 
\indent On the other hand, in November 2021, the CDC authorized vaccination for 5-11. It would be interesting to explore where incentives have been implemented for this age group, either for children or their parents.

\newpage

\section{Bibliography}

\noindent [1] Abadie, A., 2021. Using Synthetic Controls: Feasibility, Data Requirements, and Methodological Aspects. Journal of Economic Literature, 59(2), pp.391-425.\newline

\noindent [2] Abadie, Alberto, and Javier Gardeazabal. 2003. "The Economic Costs of Conflict: A Case Study of the Basque Country ." American Economic Review, 93 (1): 113-132.\newline

\noindent [3] Abadie, Alberto, Alexis Diamond and Jens Hainmueller. "Synth: An R Package for Synthetic Control Methods in Comparative Case Studies." Journal of Statistical Software, June 2011, Volume 42, Issue 13, p.1-17.
Version: Final published version.\newline

\noindent [4] Abadie, A., Diamond, A. and Hainmueller, J., 2010. Synthetic Control Methods for Comparative Case Studies: Estimating the Effect of California’s Tobacco Control Program. Journal of the American Statistical Association, 105(490), pp.493-505.\newline

\noindent [5] Brehm, M., Brehm, P. and Saavedra, M., 2021. The Ohio Vaccine Lottery and Starting Vaccination Rates. SSRN Electronic Journal.\newline

\noindent [6] Bruckhaus, A.A., Abedi, A., Salehi, S. et al. COVID-19 Vaccination Dynamics in the US: Coverage Velocity and Carrying Capacity Based on Socio-demographic Vulnerability Indices in California. J Immigrant Minority Health 24, 18–30 (2022)\newline

\noindent [7] Millar JA, Dao HDN, Stefopulos ME,
Estevam CG, Fagan-Garcia K, Taft DH, et al. (2021)
Risk factors for increased COVID-19 case-fatality in
the United States: A county-level analysis during
the first wave. PLoS ONE 16(10): e0258308.\newline

\noindent [8] American Communities Project. 2022. Home - American Communities Project. [online] Available at: <https://www.americancommunities.org/> [Accessed 29 May 2022].

\end{document}